\documentclass[aps,pre,preprint,floatfix]{revtex4}

\usepackage{latexsym}
\usepackage[usenames]{color}
\usepackage{amsthm}
\usepackage{hyperref}

\usepackage{amsmath}    % need for subequations
\usepackage{amsfonts}  %note how statements can be commented out
\usepackage{amssymb}
\usepackage{graphicx}
\usepackage{slashed}
\usepackage{fouridx}

\theoremstyle{plain} 
\theoremstyle{plain} 
\theoremstyle{plain}  
\theoremstyle{plain} 
\theoremstyle{plain}  
\theoremstyle{remark} 
\theoremstyle{plain} 
\theoremstyle{remark}

\newcommand\CROSS[1]{%
  \hbox{%
    \vbox{
      \hrule
      \kern2.5pt
      \hbox{$#1$\,\strut}
    }%
  \vrule
  }\mskip\thickmuskip
}

\begin{document}
\begin{center}
\Large{\textbf{Nonlinear embeddings: \\ Applications to analysis, fractals \\ and polynomial root finding}}\\ 
~\\

\large{Vladimir Garc\'{\i}a-Morales}\\

\normalsize{}
~\\
%Institute for Advanced Study -  Technische Universit\"{a}t M\"{u}nchen,\\ Lichtenbergstr. 2a, D-85748 Garching, Germany \\

Departament de Termodin\`amica, Universitat de Val\`encia, \\ E-46100 Burjassot, Spain
\\ garmovla@uv.es
\end{center}
\small{We introduce $\mathcal{B}_{\kappa}$-embeddings, nonlinear mathematical structures that connect, through smooth paths parameterized by $\kappa$, a finite or denumerable set of objects at $\kappa=0$ (e.g. numbers, functions, vectors, coefficients of a generating function...) to their ordinary sum at $\kappa \to \infty$. We show that $\mathcal{B}_{\kappa}$-embeddings can be used to design nonlinear irreversible processes through this connection. A number of examples of increasing complexity are worked out to illustrate the possibilities uncovered by this concept. These include not only smooth functions but also fractals on the real line and on the complex plane. As an application, we use $\mathcal{B}_{\kappa}$-embeddings to formulate a robust method for finding all roots of a univariate polynomial without factorizing or deflating the polynomial. We illustrate this method by finding all roots of a  polynomial of 19th degree. 
}
\noindent  
~\\

%\noindent complexity; predictability; time series; cellular automata; Moufang loops
\pagebreak

%\small{ \\
%%vmorales@ph.tum.de}

%A general mathematical method is presented for the systematic construction of coupled map lattices (CMLs) out of deterministic cellular automata (CAs). The entire CA rule space is addressed by means of a universal map for CAs that we have recently derived and that is not dependent on any freely adjustable parameters. The CMLs thus constructed  are termed real-valued deterministic cellular automata (RDCA) and encompass all deterministic CAs in rule space in the asymptotic limit $\kappa \to 0$ of a continuous parameter $\kappa$. Thus, RDCAs generalize CAs in such a way that they constitute CMLs when $\kappa$ is finite and nonvanishing. In the limit $\kappa \to \infty$ all RDCAs are shown to exhibit a global homogeneous fixed-point that attracts all initial conditions. A new bifurcation is discovered for RDCAs and its location is exactly determined from the linear stability analysis of the global quiescent state. In this bifurcation, fuzziness gradually begins to intrude in a purely deterministic CA-like dynamics. The mathematical method presented allows to get insight in some highly nontrivial behavior found after the bifurcation

\section{Introduction}

The question of how a composite system can be separated into parts or, vice versa, how separate, disjoint parts can be put together to form a composite system, is of general scientific interest. This question underlies the traditional distinction between linear and nonlinear systems and is manifest in the far-reaching \emph{superposition principle} obeyed by linear systems and that highlights the importance of the \emph{linear combinations} of the parts. The superimposed objects can be numbers, functions, vectors, time-varying signals, etc. The superposition principle is of fundamental importance in quantum physics: Since the Schr\"odinger equation is linear in the wave function, any of its solutions can be expressed as a linear combination (superposition) of quantum states. Mathematically, linear systems are easier to analyze because of this principle and we can wonder wether it can be generalized to nonlinear systems as well. This question is of general physical interest not only in view of the ubiquity of nonlinear systems in nature, but also, on a fundamental level because of the inherent nonlinearity of Einstein's field equations of General Relativity \cite{MTW}.

In a previous recent article \cite{CHAOSOLFRAC}, we have provided a general method, called a $p\lambda n$ decomposition, by which any nonlinear arbitrary function can be split in the linear combination of a finite set of functions which are, generally, fractal objects. Thus, the method can be regarded as a generalization of the superposition principle to any arbitrary function (linear or nonlinear). In this article, we tackle the following related question: How the parts of any decomposition can be continuously assembled together to form the composite system? To answer this question we introduce in this article the concept of nonlinear $\mathcal{B}_{\kappa}$-embeddings showing how, generally, the process of continuously assembling the parts to form a whole involves an \emph{irreversible process}. 

$\mathcal{B}_{\kappa}$-embeddings are so-called because they depend on a nonlinear function $\mathcal{B}_{\kappa}$ that we have recently introduced in a previous work \cite{JPHYSA} in promoting cellular automata \cite{Wolfram, Wolfram2, Chua1, Ilachinski, Adamatzky, VGM1, VGM2, VGM3} to coupled map lattices \cite{Kaneko1, Kapral, KanekoCrutch, KanekoBOOK}. Indeed, as shown in \cite{JPHYSA} the $\mathcal{B}_{\kappa}$-function allows a fully discrete local phase space to be embedded in a continuous one. The embeddings are governed by a parameter $\kappa$ so that, as $\kappa \to 0$ each independent part entering in the embedding is obtained. As $\kappa \to \infty$ the independence of the parts is lost and these merge with their ordinary sum.

The outline of this article is as follows. In Section \ref{bka} we briefly discuss the $\mathcal{B}$ and the $\mathcal{B}_{\kappa}$ functions and the main properties and easily derived concepts that we shall need. In Section \ref{core}, which is the core of the manuscript, the $\mathcal{B}_{\kappa}$-embedding is presented. After discussing its elementary properties, we proceed to give several examples while discussing further related concepts (some ideas related to the irreversibility of the embeddings, when viewing the parameter $\kappa$ as time, to symmetry considerations, and to possible links with nonlinear partial differential equations). We show how the embeddings can be used in analysis, to describe how a bunch of different surfaces can merge together by means of a continuous process. We apply the embeddings to the $p\lambda n$ fractal decomposition showing how the embeddings are able to continuosly carry each of the fractal objects in the decomposition at $\kappa \to 0$ to their ordinary sum at $\kappa\to \infty$. Finally, in Section \ref{apli}, we present an application of $\mathcal{B}_{\kappa}$-embeddings to one of the most important problems of numerical analysis: the solution of algebraic equations. The numerical method presented yields the location of the roots in the complex plane of an arbitrary univariate polynomial all at once.  Our method is not intended to compete with previous existing methods (as the most remarkable ones described in \cite{SW95}) and no comparison with them is provided. Rather, it is intended as an specific application of the main concept, $\mathcal{B}_{\kappa}$ embeddings, formulated in this article.

\section{The $\mathcal{B}$-function and its $\kappa$-deformation} \label{bka}

For $x, y \in \mathbb{R}$ the $\mathcal{B}$-function \cite{VGM1} is defined as
\begin{equation}
\mathcal{B}(x,y)\equiv \frac{1}{2}\left(\frac{x+y}{|x+y|}-\frac{x-y}{|x-y|}\right) \label{d1}
\end{equation}
and returns $\text{sign}(y)$ if $-y < x < y$, zero if $|x|>y$ and $\text{sign}(y)/2$ if $|x|=y$. The $\mathcal{B}$-function has the form of a rectangular function whose thickness is controlled by the value of $y$. Clearly, if $x=n-k$ is an integer and $0< y \le 1/2$, the $\mathcal{B}$-function behaves as a Kronecker delta $\delta_{nk}$, i.e.
\begin{equation}
{\displaystyle \delta_{nk}=\mathcal{B}\left(n-k,\frac{1}{2}\right)={\begin{cases} 1&{\text{if }}n=k\\0&{\text{if }}n\ne k\end{cases}}\,\!} 
\end{equation}
%For $x$ and $y$ real, we have (in the sense of distribution theory)
% \begin{eqnarray}
%\frac{\partial \mathcal{B}(x,y)}{\partial x} &=& \delta(x+y)-\delta(x-y) \label{deltabox1} \\
%\frac{\partial \mathcal{B}(x,y)}{\partial y} &=& \delta(x+y)+\delta(x-y) \label{deltabox2}
%\end{eqnarray}
%where $\delta(x)$ is the Dirac delta. Hence
%\begin{equation}
%\delta(x-y)=\frac{1}{2}\left(\frac{\partial \mathcal{B}(x,y)}{\partial y}-\frac{\partial \mathcal{B}(x,y)}{\partial x}\right)
%\end{equation}
%We also have
%\begin{equation}
%y=\frac{1}{2}\int_{-\infty}^{\infty}\mathcal{B}(x,y)dx 
%\end{equation}
%
%Note also that 
%\begin{equation}
%\mathcal{B}(x,y)=\frac{1}{2}\left(\mathcal{B}(0,x+y)-\mathcal{B}(0,x-y)\right)
%\end{equation}
%and that the maximum and minimum of $x$ and $y$ are respectively given by
%\begin{eqnarray}
%\max(x,y)&=&\frac{x+y}{2}+\frac{x-y}{2}\mathcal{B}(0,x-y) \label{maxi} \\
%\min(x,y)&=&\frac{x+y}{2}-\frac{x-y}{2}\mathcal{B}(0,x-y) \label{mini} 
%\end{eqnarray}
%Therefore
%\begin{equation}
%\mathcal{B}(x,y)=\frac{\max(x,-y)-\max(x,y)}{x+y}
%\end{equation}
%\frac{1}{2}\left(\mathcal{B}(0,x+y)-\mathcal{B}(0,x-y)\right)

The $\mathcal{B}$-function allows the straightforward construction of characteristic functions  of sets and geometric objects. One example of a set on a continuous domain is the disk, with one parameter (the radius $R$) defined as
\begin{equation}
\text{Disk}(x,y; R) \equiv \mathcal{B}(\sqrt{x^{2}+y^{2}}, R) \label{disk}
\end{equation}
This function returns 1 if the point $(x,y)$ belongs to the disk, $1/2$ if it belongs to the boundary, and zero otherwise.

We define the $\mathcal{B}_{\kappa}$-function as \cite{JPHYSA}
\begin{equation}
\mathcal{B}_{\kappa}(x,y)\equiv \frac{1}{2}\left[ 
\tanh\left(\frac{x+y}{\kappa} \right)-\tanh\left(\frac{x-y}{\kappa} \right)
\right] \label{bkappa}
\end{equation}
where $\kappa$ is a real parameter. We note the following limits for any $x$, $y$ finite and for $y\ne 0$, as proved in \cite{JPHYSA}
\begin{eqnarray}
\lim_{\kappa \to 0}\mathcal{B}_{\kappa}(x,y)&=&\mathcal{B}(x,y) \label{kronlim} \\
\lim_{\kappa \to \infty}\frac{\kappa \mathcal{B}_{\kappa}(x,y)}{y}&=&1 \label{infinlim}
\end{eqnarray}
These properties come from observing the dominant behavior of the asymptotic regimes \cite{JPHYSA}. 

The $\mathcal{B}_{\kappa}$-function, Eq. (\ref{bkappa}), is to be viewed as a smooth version of the $\mathcal{B}$-function, Eq. (\ref{d1}) which, in turn, is a suitable representation of the Kronecker delta.  The deformation parameter $\kappa$ governs the slope of the $\mathcal{B}_{\kappa}$-function at $\pm y$ from infinitely steep (at $\kappa=0$) to flat (as $\kappa \to \infty$). The $\mathcal{B}$-function is regained as the proper limit $\kappa \to 0$ of the $\mathcal{B}_{\kappa}$-function. Note that while the $\mathcal{B}$-function is compactly supported, the $\mathcal{B}_{\kappa}$-function is not, and this property is helpful to define fuzzy and broadly $\kappa$-deformed sets. For example, the disk defined in Eq. (\ref{disk}) can be transformed in a smoothed disk as follows
\begin{equation}
\text{Disk}_{\kappa}(x,y; R) \equiv \mathcal{B}_{\kappa}(\sqrt{x^{2}+y^{2}}, R) \label{disk0}
\end{equation}
In the limit $\kappa \to 0$ this expression becomes equal to $\text{Disk}_{0}(x,y; R)= \mathcal{B}(\sqrt{x^{2}+y^{2}}, R)$ which is equal to Eq. (\ref{disk}). From Eq. (\ref{infinlim}) we have $\text{Disk}_{\infty}(x,y; R)=0$ and that the asymptotic behavior in the limit $\kappa \to \infty$ is $\text{Disk}_{\kappa}(x,y; R) \sim R/\kappa$. We can parametrize $\kappa=\kappa(t)$ in terms of a time variable $t$ so that $\kappa(0)=0$ and $\kappa(\infty)=\infty$, for example as $\kappa(t)=e^{t}-1$. Thus, $\text{Disk}_{\kappa(t)}(x,y; R)$ decays to zero in time as $\sim Re^{-t}$.

%and, hence, it is not a mollifier, despite of being a smooth version of the $\mathcal{B}$-function. 

\section{Formal definition of a $\mathcal{B}_{\kappa}$-embedding, properties and examples}
\label{core}

\subsection{Definition} \label{BkE}

Let $\{f_{n \in \mathbb{Z}} \}$ be an arbitrary sequence of complex numbers or functions of an arbitrary number of variables for which ordinary multiplication by a real scalar and ordinary addition of any finite subset of terms in the sequence are well defined. We call each $f_{n}$ a \emph{part at zero}. Let $f\equiv \sum_{n\in \mathbb{Z}}f_{n}$, that we call the \emph{part at infinity}, denote the sum of all terms in the sequence. Let $\kappa \in \mathbb{R}$ be a continuous parameter, $\kappa \in [0,\infty)$. A $\mathcal{B}_{\kappa}$-embedding, $\mathcal{R}_{f}(n,\kappa)$ is defined as
\begin{equation}
\mathcal{R}_{f}(n,\kappa)\equiv (1+2\kappa)\sum_{j \in \mathbb{Z}}f_{j}\mathcal{B}_{\kappa}\left(n-j,\frac{1}{2} \right)
\end{equation}
%where
%\begin{equation}
%\mathcal{B}_{\kappa}(x,y)\equiv \frac{1}{2}\left[ 
%\tanh\left(\frac{x+y}{\kappa} \right)-\tanh\left(\frac{x-y}{\kappa} \right)
%\right] \label{bkappa}
%\end{equation}
for any $n\in \mathbb{Z}$. Thus, a $\mathcal{B}_{\kappa}$-embedding explicitly depends on the two variables $n$ and $\kappa$ (and on any other additional variable on which each separate $f_{n}$ may depend).

The main properties of the $\mathcal{B}_{\kappa}$-embedding are a simple consequence of Eqs. (\ref{kronlim}) and (\ref{infinlim}) above
\begin{eqnarray}
\mathcal{R}_{f}(n,0)&\equiv&\lim_{\kappa \to 0}\mathcal{R}_{f}(n, \kappa)=f_{n} \label{identity} \\
\mathcal{R}_{f}(n, \infty)&\equiv& \lim_{\kappa \to \infty}\mathcal{R}_{f}(n, \kappa)=\sum_{j \in \mathbb{Z}}f_{j}=f \equiv \mathcal{R}_{f}(\infty) \label{infinity}
\end{eqnarray}
These limits are easily proved from those of the $\mathcal{B}_{\kappa}$-function and justify calling each $f_{n}=\mathcal{R}_{f}(n,0)$ a \emph{part at zero} and $f=\mathcal{R}_{f}(\infty)$ the \emph{part at infinity}, 
\begin{eqnarray}
\mathcal{R}_{f}(n,0)&\equiv& \lim _{\kappa\to 0}\mathcal{R}_{f}(n,\kappa) =\lim _{\kappa\to 0}(1+2\kappa)\sum_{j \in \mathbb{Z}}f_{j}\mathcal{B}_{\kappa}\left(n-j,\frac{1}{2} \right) \nonumber \\
&=&\sum_{j \in \mathbb{Z}}f_{j}\delta_{nj}=f_{n}  \label{propi2} \\
\mathcal{R}_{f}(n, \infty)&\equiv& \lim _{\kappa\to \infty}\mathcal{R}_{f}(n,\kappa) = \lim _{\kappa\to \infty}(1+2\kappa)\sum_{j \in \mathbb{Z}}f_{j}\mathcal{B}_{\kappa}\left(n-j,\frac{1}{2} \right) \nonumber \\
&=&\lim _{\kappa\to \infty}\frac{1+2\kappa}{2\kappa}
\sum_{j \in \mathbb{Z}}f_{j}=\sum_{j \in \mathbb{Z}}f_{j}=f  \label{propi1}
\end{eqnarray}
%By using Leibniz' rule and Eq. (\ref{muitoder}) we can evaluate the derivative of any order of the embedding as
%\begin{equation}
%\frac{d^{m}\mathcal{R}_{f}(n,\kappa)}{d\kappa^{m}}=\sum_{j \in \mathbb{Z}}f_{j}\left[(1+2\kappa)
%\frac{d^{m}}{d\kappa^{m}}\mathcal{B}_{\kappa}\left(n-j,\frac{1}{2} \right)+2\frac{d^{m-1}}{d\kappa^{m-1}}\mathcal{B}_{\kappa}\left(n-j,\frac{1}{2} \right)\right]
%\end{equation}
%If we take $m=1$, we obtain
%\begin{equation}
%\frac{d\mathcal{R}_{f}(n,\kappa)}{d\kappa}= (1+2\kappa)\sum_{j \in \mathbb{Z}}f_{j}\left(\frac{2\mathcal{B}_{\kappa}\left(n-j,\frac{1}{2} \right)}{1+2\kappa}+\frac{d}{d\kappa}\mathcal{B}_{\kappa}\left(n-j,\frac{1}{2} \right) \right) \label{derRKE}
% \end{equation}
%where $\frac{d}{d\kappa}\mathcal{B}_{\kappa}\left(n-j,\frac{1}{2} \right)$ is easily obtained from Eq. (\ref{derbk}).

%\begin{equation}
%\frac{d}{d\kappa}\mathcal{B}_{\kappa}\left(n-j,\frac{1}{2} \right)=
%\frac{1}{2\kappa^{2}}\left[\left(n-j+\frac{1}{2}\right)\tanh^{2}\left(\frac{n-j+\frac{1}{2}}{\kappa} \right)-\left(n-j-\frac{1}{2}\right)\tanh^{2}\left(\frac{n-j-\frac{1}{2}}{\kappa} \right)-1
%\right] \label{derbk2}
%\end{equation}

A $\mathcal{B}_{\kappa}$-embedding continuously maps a set of parts at zero to the part at infinity. This constitutes the defining feature of a $\mathcal{B}_{\kappa}$-embedding.  The crucial fact is the following: \emph{when $\kappa \to \infty$ the embedding takes the same value $\mathcal{R}_{f}(\infty)=\sum_{j}f_{j}$, independently of $n$; for $\kappa$ finite, the embedding has different branches labelled by $n$ on which the parts at zero become accurately approximate for $\kappa$ nonvanishing but sufficiently small; when $\kappa=0$, the embedding yields exactly any of the parts at zero.} This fact is linked to the following observation: \emph{Because of the fast saturation of the $\mathcal{B}$-function, the limiting values, as $\kappa \to \infty$ or $\kappa \to 0$ are well approximated for $\kappa$ finite sufficiently large or sufficiently small, respectively}.

%\begin{eqnarray}
%\frac{d^{n}}{d\kappa^{n}}\mathcal{B}_{\kappa}(x,y)&=&\frac{2^{n-1}(x+y)^{n}e^{2(x+y)/\kappa}}{\kappa^{n+1}(1+e^{2(x+y)/\kappa})^{n+1}}\sum_{k=0}^{n-1} \left\langle {n \atop k} \right\rangle (-1)^{k+1}e^{2k(x+y)/\kappa} \nonumber \\
%&&-\frac{2^{n-1}(x-y)^{n}e^{2(x-y)/\kappa}}{\kappa^{n+1}(1+e^{2(x-y)/\kappa})^{n+1}}\sum_{k=0}^{n-1} \left\langle {n \atop k} \right\rangle (-1)^{k+1}e^{2k(x-y)/\kappa}
%\end{eqnarray}

%To the curve defined parametrically in terms of $\kappa$ with coordinates $(\kappa, \mathcal{R}_{f}(\kappa))$ thus corresponds a gradient curve with coordinates $(\kappa, d\mathcal{R}_{f}/d\kappa)$ and a \textbf{projective dual curve} \cite{Arnold} given by 
%\begin{equation}
%\left(\frac{d\mathcal{R}_{f}}{d\kappa},\ \kappa \frac{d\mathcal{R}_{f}}{d\kappa}-\mathcal{R}_{f} \right)
%\end{equation} 

As we shall see below, the robustness of the limits $\kappa \to 0$ and $\kappa \to \infty$ has important consequences when functions composed of parts with certain symmetries are considered, the embedding being a useful means of continuously breaking these symmetries. We also note that, while Eqs. (\ref{identity}) and (\ref{infinity}) are invariant under shift of the labels $j \to j'$ so that $f_{j} \to f_{j'}$, the $\mathcal{B}_{\kappa}$-embeddings themselves are not invariant under these transformations for finite non-vanishing $\kappa$. However, since the limiting values (i.e. the parts at zero and at infinity) are the same we call such embeddings \emph{equivalent} and consider any of them as a representative of the same class. If any of the parts at zero or the part at infinity are different, we say that the embeddings belong to different classes.
 
%Changing $\kappa$ in a $\mathcal{B}_{\kappa}$-embedding implies changing the support of the latter from being compact (at $\kappa=0$) to cover the whole real line (at $\kappa = \infty$). 
Of course, a single part $f_{0}$ at zero can be mapped to itself at infinity by means of the embedding
\begin{equation}
 \mathcal{R}_{f}(n,\kappa)=(1+2\kappa)f_{0}\mathcal{B}_{\kappa}\left(n, \frac{1}{2}\right)
\end{equation} 
We call such an embedding consisting of a single part connected to itself a \emph{loop}. 

Because at $\kappa \to \infty$ the embedding is insensitive to $n$ we can safely omit the effect of the label $n$ by setting all $\mathcal{B}_{\kappa}$-functions and the prefactor $(1+2\kappa)$ to unity. As we shall see, it is remarkable that this insensitivity to the value of $n$ is already present, \emph{for every practical purposes}, for $\kappa$ sufficiently large. Note, furthermore that since $\kappa$ is a continuous parameter, its increase from zero to infinity is similar to the increase of a time variable $t$ from an initial condition to the future $t>0$ and the $\mathcal{B}_{\kappa}$-embedding can be considered as the solution of a flow. This view of the embedding in terms of a dynamical system shall be fruitfully exploited below.  Note that we can also reverse the 'arrow of time' if we e.g. make the transformation $\kappa \to 1/\kappa$. Any $\mathcal{B}_{\kappa}$-embedding is reversible under this transformation \emph{for finite $\kappa$}. However, if the limit $\kappa=t \to \infty$ is strictly taken, this reversibility is lost and it is impossible to know with certainty from which specific part at zero $f_{n}$ the flow started at $\kappa=t=0$.

\subsection{Example 1: Embeddings with natural numbers as parts at zero and at infinity}

\begin{figure}
\includegraphics[width=0.75 \textwidth]{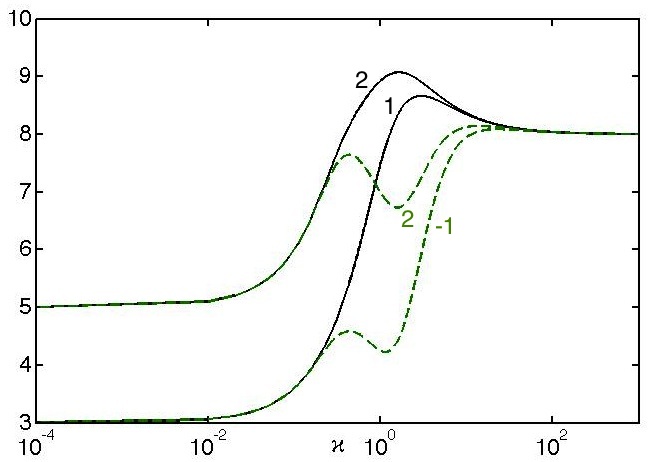}
\caption{\scriptsize{The $\mathcal{B}_{\kappa}$-embeddings given  by Eqs. (\ref{RKE1a}) (continuous curves) and (\ref{RKE1b}) (dashed curves) for the values of $n$ indicated in the figure. Since both embeddings have the same parts at zero and at infinity, they are on the same class.}} \label{pRKE}
\end{figure}

Any embedding connecting $N_{0}$ natural numbers at zero with their sum at infinity looks, when represented as a function of $\kappa$, as a twisted fork with $N_{0}$ prongs. Let us illustrate this and all above remarks with specific examples. The natural number 8 can be expressed as the sum of the naturals 3 and 5 and we can then construct a $\mathcal{B}_{\kappa}$-embedding taking  $f_{1}=3$, $f_{2}=5$ and $f_{j}=0$ $\forall j\in \mathbb{Z}, j \notin \{1,2\}$ as parts at zero. Then, 
\begin{equation}
\mathcal{R}_{f}(n,\kappa)=(1+2\kappa)\left[3\mathcal{B}_{\kappa}\left(n-1,\frac{1}{2}\right)+5\mathcal{B}_{\kappa}\left(n-2,\frac{1}{2}\right)\right] \label{RKE1a}
\end{equation}
is the resulting $\mathcal{B}_{\kappa}$-embedding. We have
\begin{eqnarray}
\mathcal{R}_{f}(1,0)&=&3 \qquad \mathcal{R}_{f}(2,0)=5 \qquad \left(\mathcal{R}_{f}(n,0)=0 \quad \forall n \notin\{1,2\} \right) \nonumber \\
\mathcal{R}_{f}(\infty)&=&8  \nonumber
\end{eqnarray}
For all $\kappa$ finite and nonvanishing, $\mathcal{R}_{f}(n,\kappa)$ given by Eq. (\ref{RKE1a}) nonlinearly connects the two parts at zero and the one at infinity. 
The labels $j \in \{1,2\}$ can be shifted to any other integer values and the resulting embedding is in the same class. If we, e.g., take $j \in \{-1,2\}$ we have, instead of Eq. (\ref{RKE1a}) the embedding
\begin{equation}
\mathcal{R}_{f}(n,\kappa)=(1+2\kappa)\left[3\mathcal{B}_{\kappa}\left(n+1,\frac{1}{2}\right)+5\mathcal{B}_{\kappa}\left(n-2,\frac{1}{2}\right)\right] \label{RKE1b}
\end{equation}
which is equivalent to the embedding in Eq. (\ref{RKE1a}). Both embeddings are represented in Fig. \ref{pRKE} for the values of $n$ indicated in the figure. As indicated above, the embeddings look like severely twisted forks with two prongs (in this case the number of parts at zero is $N_{0}=2$). Each prong correspond to a different branch of the embedding labelled by $n$. We see that the asymptotic values at $\kappa=0$ and $\kappa=\infty$ are the same for both embeddings and that their respective branches can be continuously deformed from one embedding to the other. The branches only differ significantly at intermediate $\kappa$ values where, because of the nonlinear character of the $\mathcal{B}_{\kappa}$ function, they exhibit complicated behavior, displaying extremal points. The parts and zero and at infinity are accurately approximated for finite non-vanishing $\kappa$ values on wide regions of the parameter space (note the logarithmic scale of the axis in the figure), because of the saturation properties of the $\mathcal{B}_{\kappa}$-function.

As mentioned in Section \ref{BkE}, the above $\mathcal{B}_{\kappa}$-embedding can be used to model a simple irreversible process. Indeed, we now show the following: \emph{If the parts at zero and at infinity are constrained to be natural numbers, the information lost in mapping through an embedding a part at zero with value $M\in \mathbb{N}$ to the part at infinity with value $N \in \mathbb{N}$  can be quantified in terms of a non-negative entropy change }
\begin{equation}
\Delta S_{M \to N}=\ln \frac{p(N)}{p(M)} \ge 0 \label{entroincre}
\end{equation} 
\emph{where $p(N)$ and $p(M)$ are the number of unrestricted integer partitions of $N$ and $M$ respectively.} The latter are exactly given by the Hardy-Ramanujan-Rademacher formula (see \cite{AndrewsParti}, Chapter 5) as  
\begin{equation}
p(N)=\frac{1}{\pi \sqrt{2}}\sum_{k=1}^{\infty}\sum_{\begin{smallmatrix}0\le h \le k-1\\ (h,k)=1\end{smallmatrix}}e^{\pi i\left(s(h,k)-\frac{2 hN}{k}\right)}k^{1/2}\left.\frac{d}{dx}
\frac{\sinh \left(\frac{\pi}{k}\left(\frac{2}{3}\left(x-\frac{1}{24}\right)\right)^{1/2}      \right)}{\left(x-\frac{1}{24}\right)^{1/2}}\right|_{x=N}  \label{HRR}
\end{equation}
where $(h,k)=1$ means that $h$ and $k$ are coprime and $s(h,k)$ is the Dedekind sum given by
\begin{equation}
s(h,k)=\sum_{\mu=1}^{k-1}\left(\frac{\mu}{k}-\left \lfloor \frac{\mu}{k} \right \rfloor-\frac{1}{2}\right)\left(\frac{h\mu}{k}-\left \lfloor \frac{h\mu}{k} \right \rfloor-\frac{1}{2}\right) 
\end{equation}
with $\left \lfloor \ldots \right \rfloor$ denoting the floor function (i.e. the closest integer obtained from rounding to minus infinity).

To substantiate Eq. (\ref{entroincre}), note that if the part at infinity has value $M\in \mathbb{N}$, there are $p(M)$ different ways of choosing natural numbers so that their sum is equal to $M$, where $p(M)$ is the total number of unrestricted partitions of $M$. This number is given by Eq. (\ref{HRR}) for any $M \in \mathbb{N}$ \cite{AndrewsParti}. For example, since $4$ can be written as a sum of natural numbers in five different ways as $4$, $1+3$, $1+1+2$, $2+2$, $1+1+1+1$ we have $p(4)=5$. These possible partitions coincide with the total number of classes for a $\mathcal{B}_{\kappa}$-embeddings connecting the parts at zero, taken from these integer partitions, to the part at infinity, with value $M$. We note that $p(1)=1$ and, hence, the only possible embedding with natural-valued part at zero having part at infinity equal to 1 is a loop. Thus, if the part at infinity has value $M$, if we take a possible embedding \emph{at random} with parts at zero constrained to be natural numbers, the probability that the embedding belongs to a specific class is $1/p(M)$. Therefore, we can formally associate a (dimensionless) `Boltzmann entropy' to this probability as
\begin{equation}
S(M)=-\ln \frac{1}{p(M)}
\end{equation}
If we now consider an embedding with $M$ as a possible part at zero and $N \in \mathbb{N}$ as the part at infinity, we have, since $p(N)$ is a strictly increasing function of $N$, $p(N) \ge p(M)$ where the equality only holds if $M=N$ (i.e. if the embedding is a loop). Therefore, 
\begin{equation}
\Delta S_{M \to N} \equiv S(N)-S(M)=-\ln \frac{1}{p(N)}+\ln \frac{1}{p(M)}=\ln \frac{p(N)}{p(M)} \ge 0
\end{equation}
The number of integer partitions $p(N)$ given by Eq. (\ref{HRR}) strongly increases with $N$. One has, e.g., $p(10)=42$, $p(100)=190569292$, $p(200)=3972999029388$, etc.

We can now evaluate the positive entropy increase in going from the parts at zero 3 and 5 to the part at infinity 8 in the $\mathcal{B}_{\kappa}$-embedding given by Eq. (\ref{RKE1a}). We have
\begin{eqnarray}
\Delta S_{3 \to 8}&=& \ln \frac{p(8)}{p(3)}=\ln \frac{22}{3}=1.99243\ldots \nonumber \\
\Delta S_{5 \to 8}&=& \ln \frac{p(8)}{p(5)}=\ln \frac{22}{7}=1.14513\ldots
\end{eqnarray}
for the branches starting at zero on parts 3 and 5, respectively. If we let $\kappa=t$ where $t$ is time, the flows given by the embeddings in Eqs. (\ref{RKE1a}) and (\ref{RKE1b}) represent possible irreversible evolutions from two different initial equilibria (parts at zero) to a final equilibrium state (part at infinity) with positive `entropy changes' as calculated above, thus satisfying a `second law of thermodynamics'. Note that for a sufficiently long time, as the part at infinity is approached, it is impossible, in practice, to exactly revert the trajectories and the information of the initial condition is lost.

To better understand this loss of information, we note that in Fig. \ref{pRKE} the parts at zero can be put in a vector $(3,5)$. At infinity, this vector is lost and we have, indeed, a natural scalar 8. All detailed information from the components of the vector (and the dimensionality of the vector space at zero) is lost. It is then impossible to know at infinity, whether we started at zero from $(3,5)$, $(6,2)$, $(4,4)$, $(1,1,1,1,4)$, etc. \emph{If the limit is not strictly taken ($\kappa$ finite) we can always revert the embedding and know the initial condition by going backward (by decreasing $\kappa$)} on the corresponding branch of the embedding. This situation is analogous to what one encounters in thermodynamics (where there is an arrow of time and, hence, irreversibility) as opposed to classical mechanics (which is time-reversible) \cite{Annals, Lebowitz}. In thermodynamics, a scalar function (a thermodynamic potential) and the knowledge of few macroscopic variables suffices to characterize the state of the macroscopic system of $N$ particles (where $N$ is enormous). The detailed state of the latter consists indeed on the values of $6N$ degrees of freedom, $3N$ for the momenta of the particles and $3N$ for their positions, the behavior of the $6N$ degrees  of freedom being described by classical mechanics. Thus, passing from a description governed by classical mechanics to the one described by thermodynamics is analogous to following a $\mathcal{B}_{\kappa}$-embedding from the parts at zero to the part at infinity ('thermalization').

It is to be noted that for every natural number $N$ there are, besides the trivial partitions $N=\sum_{k=1}^{N}1$ and $N=0+N$, many other partitions that are easy to construct. Note that any natural number $N$ can be expressed in radix (base) $p \in \mathbb{N}$, $p\ge 2$ in a unique way as \cite{CHAOSOLFRAC, Andrews,QUANTUM, PHYSAFRAC, semipredo} 
\begin{equation}
N=\sum_{k=0}^{\left \lfloor \log_{p} N \right \rfloor}p^{k}\mathbf{d}_{p}(k,N) \label{naturad}
\end{equation}
where we have introduced the digit function \cite{CHAOSOLFRAC}
\begin{equation}
\mathbf{d}_{p}(k,x) = \left \lfloor \frac{x}{p^{k}} \right \rfloor-p\left \lfloor \frac{x}{p^{k+1}} \right \rfloor   \label{cucuAreal}
\end{equation}
for $x \in \mathbb{R}$ and $k \in \mathbb{Z}$. It is clear then that from Eq. (\ref{naturad}) a total number of $2^{\left \lfloor \log_{p} N \right \rfloor+1}$ embeddings (not all necessarily different) can be constructed for each choice of the radix $p$ by making $N_{0}$ parts at zero ($1 \le N_{0} \le \left \lfloor \log_{p} N \right \rfloor+1$) out of partial sums of the terms in in Eq. (\ref{naturad}). 

Any natural number $N$ can also be expressed as the product of $N_{0}$ prime factors $p_{j}$ (not necessarily distinct) in a unique way
\begin{equation}
N=\prod_{j=1}^{N_{0}}p_{j} \label{multin}
\end{equation}
By taking the logarithm to both sides
\begin{equation}
\ln N=\sum_{j=1}^{N_{0}}\ln p_{j}
\end{equation}
we observe that the $\mathcal{B}_{\kappa}$-embedding
\begin{equation}
\mathcal{R}_{N}(n,\kappa)=(1+2\kappa)\sum_{j=1}^{N_{0}}\mathcal{B}_{\kappa}\left(n-j,\frac{1}{2} \right)\ln p_{j}
\end{equation}
sends each part $\ln p_{n}$ at zero to $\ln N$ at infinity. We further note that
\begin{equation}
\mathcal{R}_{N}(n,\kappa)=\sum_{j=1}^{N_{0}}\ln p_{j}^{(1+2\kappa)\mathcal{B}_{\kappa}\left(n-j,\frac{1}{2} \right)}=\ln \left(\prod_{j=1}^{N_{0}} p_{j}^{(1+2\kappa)\mathcal{B}_{\kappa}\left(n-j,\frac{1}{2} \right)}\right)
\end{equation}
Thus, the function
\begin{equation}
e^{\mathcal{R}_{N}(n,\kappa)}=\prod_{j=1}^{N_{0}} p_{j}^{(1+2\kappa)\mathcal{B}_{\kappa}\left(n-j,\frac{1}{2} \right)} \label{multemb}
\end{equation}
is a multiplicative embedding that continuously sends each prime factor $p_{n}$ of $N$ at zero to $N$ at infinity, since we have
\begin{eqnarray}
e^{\mathcal{R}_{N}(n,0)}&=&p_{n} \nonumber \\
e^{\mathcal{R}_{N}(n,\infty)}&=&N \label{multend}
\end{eqnarray}
Taking the prime factors as parts at zero of the multiplicative embedding means again considering a specific equivalence class in the space of embeddings. If $N$ is prime, the only multiplicative embedding with natural parts at zero is the trivial one (a loop). Let now $m(N)$ denote the number of multiplicative partitions of $N$ and let $M$ be a factor of $N$. We then, have, again
\begin{equation}
\Delta S_{M \to N}=\ln \frac{m(N)}{m(M)} \ge 0 \label{entromultincre}
\end{equation} 
for the multiplicative embedding above. Thus, following the multiplicative embedding from $\kappa=0$ to $\infty$ can also be seen as the unfolding of an irreversible process.

\subsection{Example 2: Embeddings having functions of real or complex variables as parts}

The parts of an embedding can freely be any mathematical structures for which ordinary addition, as well as multiplication by an scalar, are well defined. They can also be functions, vectors, matrices, etc. For example, let us consider real-valued functions of a real variable $x$. The following embedding
\begin{equation}
\mathcal{R}_{f(x)}(n,\kappa)=(1+2\kappa)\left[\left(5\sin x\right)\mathcal{B}_{\kappa}\left(n-1, \frac{1}{2}\right)+\left(17\cos x+68-\frac{x^{2}}{25}\right)\mathcal{B}_{\kappa}\left(n-2, \frac{1}{2}\right)\right] \label{RKE2}
\end{equation}
has parts at zero $f_{1}(x)=5\sin x$ and $f_{2}(x)=17\cos x+68-\frac{x^{2}}{25}$. At infinity,
\begin{equation}
f(x)=\mathcal{R}_{f(x)}(\infty)=5\sin x+17\cos x+68-\frac{x^{2}}{25} \label{RKE2b}
\end{equation}

\begin{figure}
\includegraphics[width=0.75 \textwidth]{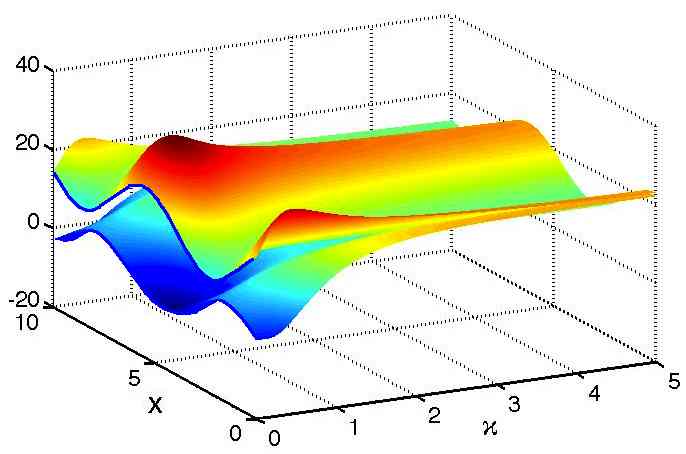}
\caption{\scriptsize{The $\mathcal{B}_{\kappa}$-embedding given by Eq. (\ref{RKE2}). The two parts at $\kappa=0$, $f_{1}(x)=5\sin x$ and $f_{2}(x)=17\cos x+68-\frac{x^{2}}{25}$ are the blue curves. The embedding continuously maps these curves to the part at infinity given by Eq. (\ref{RKE2b}).}} \label{contiRK}
\end{figure}

This embedding is represented in Fig. \ref{contiRK} as a function of $\kappa$. At infinity there is a single curve given by Eq. (\ref{RKE2b}) where the two surfaces $f_{1}(x,\kappa)=\mathcal{R}_{f(x)}(1,\kappa)$ and $f_{2}(x,\kappa)=\mathcal{R}_{f(x)}(2,\kappa)$ collapse. At $\kappa=0$ one obtains the two curves  $f_{1}(x)=5\sin x$ and $f_{2}(x)=17\cos x+68-\frac{x^{2}}{25}$, which are the parts at zero of the embedding.

Note that although one of the parts at zero $f_{1}(x)=5\sin x$ is a periodic function, the other part is not and the embedding is neither. When $\kappa$ is small but nonzero, the periodicity of the branch $\mathcal{R}_{f(x)}(1,\kappa)$ is only weakly broken by the small term
\begin{equation}
(1+2\kappa)\left(17\cos x+68-\frac{x^{2}}{25}\right)\mathcal{B}_{\kappa}\left(-1, \frac{1}{2}\right)
\end{equation}
that reflects the influence of the other branch of the embedding. In general, let each part of an embedding $f_{j}(x)$ be periodic with period $T_{j}$ so that $f_{j}(x+T_{j})=f_{j}(x)$. If all periods are integer numbers, the embedding shall also be a periodic function, with period equal to the least common multiplier of all the $T_{j}$.  

We now note the following fact. If all parts at zero $f_{j}(x)$ of an embedding are invariant under a transformation $x \to x'$ then, then the part at infinity is invariant, as well, under the same transformation. However, the contrary is not true and \emph{there may be situations were the part at infinity is invariant under a transformation but the parts at zero are not}. A simple example is the following. Let a function $f(x)$ be such that $f(x) \neq f(-x)$, i.e. a function that is not invariant under reflection $x \to -x$. Let $g(x)=f(x)+f(-x)$.
Then, the embedding
\begin{equation}
\mathcal{R}_{g(x)}(n,\kappa)=(1+2\kappa)\left[f(x)\mathcal{B}_{\kappa}\left(n-1, \frac{1}{2}\right)+f(-x)\mathcal{B}_{\kappa}\left(n-2, \frac{1}{2}\right)\right]
\end{equation}
is symmetric under reflection at infinity, since one has there $\mathcal{R}_{g(x)}(\infty)=g(x)=f(x)+f(-x)$ (and, trivially $g(x)=g(-x)$) but it is not symmetric at zero, where one has either $f(x)$ or $f(-x)$, which are not invariant under reflection, by assumption. Therefore, \emph{the part at infinity contains all symmetries which are common to all parts at zero plus additional symmetries that may not be present in the latter when taken individually}. 

%However, quite importantly, \emph{if the part at infinity is invariant under a certain transformation, the \emph{set} of parts at zero is invariant under the same transformation.}

%We have also the following fact, that links $\mathcal{B}_{\kappa}$-embeddings with bifurcation theory, a cornerstone in the analysis of systems of nonlinear differential equations of mathematical physics:   This means that the branches of an embedding can be interpreted as the branches of symmetry-breaking solutions of ordinary differential equations that bifurcate when a parameter is varied.

Note that if a certain symmetry is possessed by an individual part at zero \emph{but is not common to the other parts in the set}, that particular symmetry may also be absent at infinity. For example, for the function $f(x)$ above, such that $f(x)\ne f(-x)$ one can also construct the embedding
\begin{equation}
\mathcal{R}_{f(x)}(n,\kappa)=(1+2\kappa)\left[\frac{f(x)+f(-x)}{2}\mathcal{B}_{\kappa}\left(n-1, \frac{1}{2}\right)+\frac{f(x)-f(-x)}{2}\mathcal{B}_{\kappa}\left(n-2, \frac{1}{2}\right)\right]
\end{equation}
that connects the symmetric and antisymmetric parts of $f(x)$ at $\kappa=0$ to the function $f(x)$ at $\kappa=\infty$. If $f(x)$ is even $f(x)=f(-x)$ or odd $f(x)=-f(-x)$ this latter embedding becomes a loop.

If we want to connect a function $P(x)$ at $\kappa=0$ to another function $Q(x)$ at $\kappa=\infty$, this is achieved by means of the embedding
\begin{equation}
\mathcal{R}_{Q(x)}(n,\kappa)=(1+2\kappa)\left[P(x)\mathcal{B}_{\kappa}\left(n-1, \frac{1}{2}\right)+(Q(x)-P(x))\mathcal{B}_{\kappa}\left(n-2, \frac{1}{2}\right)\right] \label{homy1}
\end{equation} 
by following the branch with $n=1$ while increasing $\kappa$. If there are $N$ functions $P_{j}(x)$, $j=1,2,\ldots, N$ that are each to be individually mapped to $Q(x)$ at infinity, we have
\begin{equation}
\mathcal{R}_{Q(x)}(n,\kappa)=(1+2\kappa)\left[\sum_{j=1}^{N}P_{j}(x)\mathcal{B}_{\kappa}\left(n-j, \frac{1}{2}\right)+\left(Q(x)-\sum_{j=1}^{N}P_{j}(x)\right)\mathcal{B}_{\kappa}\left(n-N-1, \frac{1}{2}\right)\right] \label{homy2}
\end{equation} 
which generalizes Eq. (\ref{homy1}). Let us, for example, consider the Taylor expansion of a function that is continuous and, at least, $N$ times differentiable
\begin{equation}
f(x+h)=\sum_{j=0}^{N}\frac{h^{j}}{j!}\left.\frac{d^{j}f(y)}{dy^{j}}\right|_{y=x}+\epsilon
\end{equation}
Where $\epsilon$ is the error committed in the truncation. If we take $P_{j}(x)=\frac{h^{j}}{j!}\left.\frac{d^{j}f(y)}{dy^{j}}\right|_{y=x}$ and $Q(x)=f(x+h)$ in Eq. (\ref{homy2}) we obtain the embedding
\begin{equation}
\mathcal{R}_{f(x+h)}(n,\kappa)=(1+2\kappa)\left[\sum_{j=1}^{N}\frac{h^{j}}{j!}\left.\frac{d^{j}f(y)}{dy^{j}}\right|_{y=x}\mathcal{B}_{\kappa}\left(n-j, \frac{1}{2}\right)+\epsilon\mathcal{B}_{\kappa}\left(n-N-1, \frac{1}{2}\right)\right] \label{homy3}
\end{equation} 
The latter sends each term on the Taylor expansion $P_{n}(x)=\frac{h^{n}}{n!}\left.\frac{d^{n}f(y)}{dy^{n}}\right|_{y=x}$ at zero to $f(x+h)$ at infinity. Since $P_{0}(x)=f(x)$, following the specific branch $n=0$ of the embedding from $\kappa=0$ to $\infty$ is equivalent to applying the shift operator $\mathbf{T}^{h}f(x)=f(x+h)$.

We can also consider several values of $\kappa$ that depend on $n$ and which are  monotonously increasing function of time $\kappa_{n}(t)$ ($\kappa_{n}(0)=0$, $\kappa_{n}(\infty)=\infty$) and make the part at infinity also time dependent. Thus Eq. (\ref{homy2}) can be generalized to
\begin{eqnarray}
\mathcal{R}_{Q(x,t)}(n,\kappa_{1}(t),\ldots, \kappa_{N+1}(t))&=&\sum_{j=1}^{N}P_{j}(x)(1+2\kappa_{j}(t))\mathcal{B}_{\kappa_{j}(t)}\left(n-j, \frac{1}{2}\right) \label{homy3} \\
&& +\left(Q(x,t)-\sum_{j=1}^{N}P_{j}(x)\right)(1+2\kappa_{N+1}(t))\mathcal{B}_{\kappa_{N+1}(t)}\left(n-N-1, \frac{1}{2}\right) \nonumber
\end{eqnarray} 
Thus, at $t=0$ we would have any of the $P_{n}(x)$ as initial condition and, after a transient, for $t$ sufficiently large, all $\mathcal{B}_{\kappa(t)}$-functions saturate and we are left with
\begin{equation}
\mathcal{R}_{Q(x,t)}(n,\kappa(t))\sim \mathcal{R}_{Q(x,t)}(n,\infty)=Q(x,t) \label{homy3b}
\end{equation} 
i.e. with Eq. (\ref{homy3}) \emph{we are able to model the same kind of behavior as with any arbitrary nonlinear partial differential equation}: starting from an initial condition, we have, after a transient, a behavior governed for $t$ large by $Q(x,t)$. This latter function can be an oscillating function of time, a standing pattern, etc. Although Eq. (\ref{homy3}) may look complicated, its interpretation is straightforward, because it is just the ordinary addition of the loops connecting each part to itself, the loops being unfolded on different time scales. Thus Eq. (\ref{homy3}) is a linear superposition of nonlinear embeddings and, although the transient behavior is highly nontrivial, the behavior at zero and at infinity is clear. 

We thus conjecture that Eq. (\ref{homy3}) can be used to find solutions of nonlinear partial differential equations as follows. Let us consider, for example, the complex Ginzburg-Landau equation \cite{KuramotoBOOK, Aranson, contemphys} on a ring of length $L$. Let $W(x,t)$ denote the dynamical variable (complex order parameter) whose spatiotemporal evolution we want to know. Since it is necessary to satisfy periodic boundary conditions, it is clear that $W(x,t)$ must have the following form
\begin{equation}
W(x,t)=\sum_{k=-\infty}^{\infty}A_{k}(t)e^{i2\pi k x/ L}
\end{equation}
in order to warrant that $W(x+L,t)=W(x,t)$ at every $t$. Suppose now that, with help of computer simulations, we find a set of $N$ initial conditions that display the same asymptotic behavior as $t \to \infty$. Since $W(x,t) \sim Q(x,t)$ in that regime, it is now clear that $Q(x,t)=\sum_{k=-\infty}^{\infty}Q_{k}(t)e^{i2\pi k x/ L}$ and we can make a Fourier analysis of such asymptotic behavior to estimate the $Q_{k}(t)$ and, thus, guess from this empirical knowledge the function $Q(x,t)$ mathematically (taking care of the observed symmetries, etc). The initial conditions would be our $P_{j}(x)$ (our known input) in Eq. (\ref{homy3}) and the asymptotic behavior so constructed can now be used as $Q(x,t)$ in Eq. (\ref{homy3}). Then, by making the ansatz $W(x,t)=\mathcal{R}_{Q(x,t)}(n,\kappa_{1}(t),\ldots, \kappa_{N+1}(t))$ in the nonlinear partial differential equation, we would expect to find conditions on the $\kappa_{j}(t)$ that constrain their behavior. If these constraints can analytically be satisfied, we would conclude that we have found $N$ exact trajectories of the nonlinear partial differential equation in question. We shall attempt to carry out this program elsewhere.

\subsection{Example 3: Embeddings having sets as parts}

\begin{figure*}
\includegraphics[width=1.0 \textwidth]{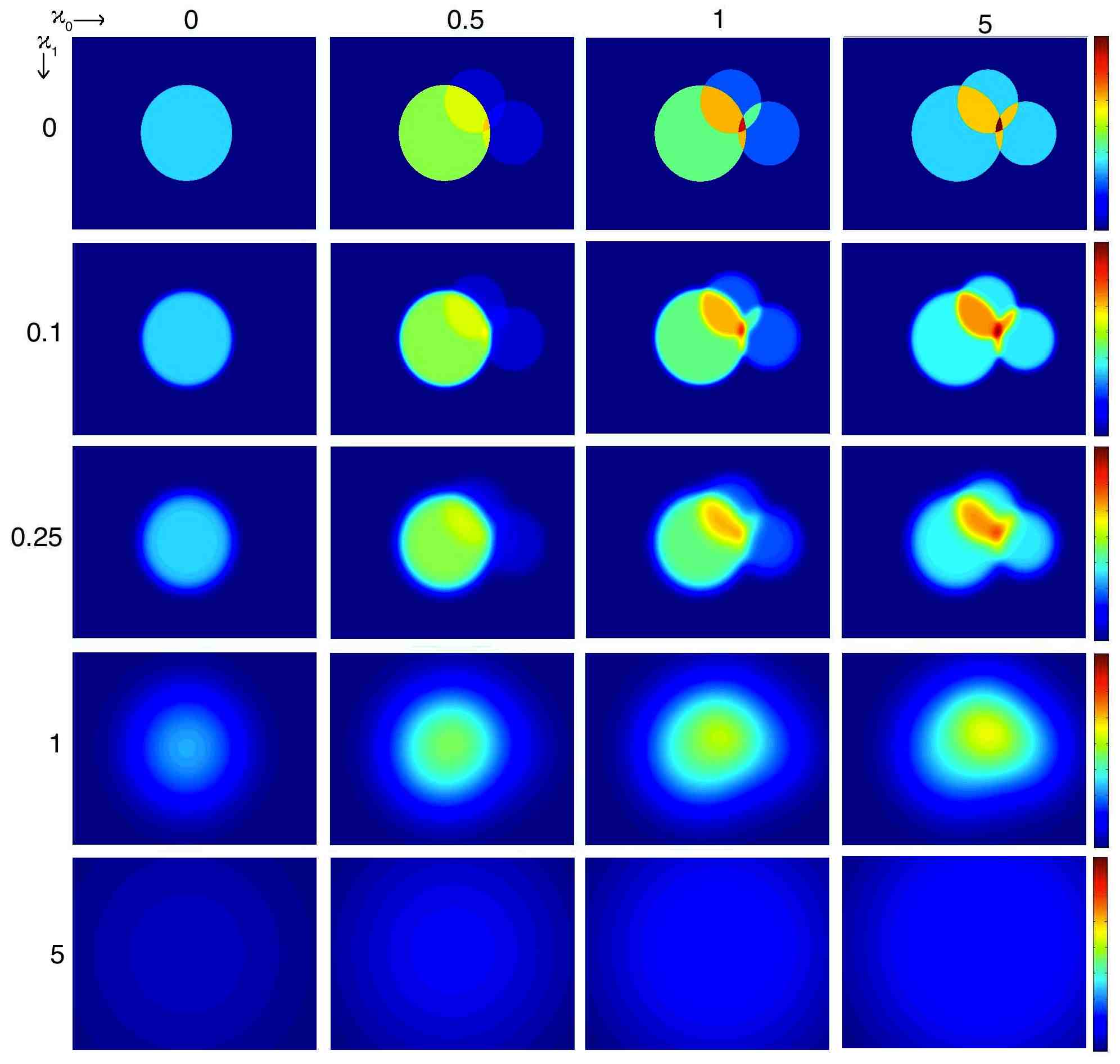}
\caption{\scriptsize{Plot of $\mathcal{R}_{S(x,y; \kappa_{1})}(1,\kappa_{0})$ obtained from Eq. (\ref{disky}) for $n=1$ and the values of $\kappa_{0}$ and $\kappa_{1}$ indicated on the headings of the columns and rows of panels. Each panel shows a region of the plane for which $x \in [-3,5]$ and $y\in [-3,3]$. The color palette ranges linearly from zero (dark blue) to 3 (dark red).}} \label{disko}
\end{figure*}

Characteristic functions of sets and geometric objects constructed with help of the $\mathcal{B}$ or the $\mathcal{B}_{\kappa}$ functions can also be embedded as parts at zero to yield their ordinary addition at infinity. Different, independent deformation parameters can also be considered. For example, the embedding 
\begin{eqnarray}
\mathcal{R}_{S(x,y; \kappa_{1})}(n,\kappa_{0})&=&(1+2\kappa_{0})\left[\text{Disk}_{\kappa_{1}}(x-3,y; 1)\mathcal{B}_{\kappa_{0}}\left(n, \frac{1}{2}\right)+\text{Disk}_{\kappa_{1}}(x-0.75,y; 1.5)\mathcal{B}_{\kappa_{0}}\left(n-1, \frac{1}{2}\right)+\right. \nonumber \\
&&\ \qquad \qquad +\left.\text{Disk}_{\kappa_{1}}(x-1.75,y-1;1)\mathcal{B}_{\kappa_{0}}\left(n-2, \frac{1}{2}\right)
\right] \label{disky}
\end{eqnarray} 
contains the superposition of three disks defined by Eq. (\ref{disk0}), centered at points $(x_{0},y_{0})=(3,0)$, $(0.75,0)$, $(1.75,1)$ and with respective radii $R=1$, $1.5$ and $1$. 
The deformation parameter $\kappa_{1}$ controls how smooth the border of each individual disk is (from sharp at $\kappa_{1} \to 0$ to smoother for $\kappa_{1}$ nonvanishing). The parameter $\kappa_{0}$ controls the embedding of the three disks individually taken as parts at zero carrying them into coexistence at infinity, where the embedding becomes equal to
\begin{equation}
S(x,y; \kappa_{1})=\text{Disk}_{\kappa_{1}}(x-3,y; 1)+\text{Disk}_{\kappa_{1}}(x-0.75,y; 1.5)+
\text{Disk}_{\kappa_{1}}(x-1.75,y-1;1) \label{sdisky}
\end{equation}
In the limit $\kappa_{1}\to 0$ the function $S(x,y; 0)$ has the appearance of a Venn diagram, and one has $S(x,y; 0)=N$ where $N$ is the number of discs to whose interior the point $(x,y)$ belongs (if $(x,y)$ belongs to the interior of $N_{0}$ discs and to the boundary of $N_{1}$ discs, one has $S(x,y; 0)=N_{0}+N_{1}/2$).

In Fig. \ref{disko} we plot $\mathcal{R}_{S(x,y; \kappa_{1})}(1,\kappa_{0})$ obtained from Eq. (\ref{disky}) for the values of $\kappa_{0}$ and $\kappa_{1}$ indicated and $n=1$. We observe that at $\kappa_{0} \to 0$, $\mathcal{R}_{S(x,y; \kappa_{1})}(1,0)=\text{Disk}_{\kappa_{1}}(x-0.75,y; 1.5)$. Then, as $\kappa_{0} \approx 5$ the infinite limit of the embedding is already excellently approximated, i.e. $\mathcal{R}_{S(x,y; \kappa_{1})}(1,5) \approx S(x,y; \kappa_{1})$, where $S(x,y; \kappa_{1})$ is given by Eq. (\ref{sdisky}). Note that we could have started as well from any other disk, by considering the corresponding branch labelled by $n$, and the asymptotic behavior on the $\kappa_{0}$ direction would be the same (provided that $\kappa_{1}$ is kept fixed constant). 

If we set $\kappa_{0}=e^{c_{0}t}-1$ and $\kappa_{1}=e^{c_{1}t}-1$, where $c_{0}$ and $c_{1}$ are constants and $t$ is time, we see that the increase in $\kappa_{1}$ is  similar to a nonlinear diffusion process blurring any initial structure while the increase in $\kappa_{0}$ (governing the embedding) is much more subtle: The initial structure is assembled to other structures in the course of time until it gets to a state $\kappa_{0} \to \infty$ where all structures coexist. Thus, certain \emph{emergent} properties of nonlinear dynamical systems can be modeled in this way. If we take $c_{1}<<c_{0}$ we can, for example, model through the change of $\kappa_{1}$ the slow aging and eventual disappearance of an organism (whose structure, starting from separate elements, is assembled faster on the direction of the increase of $\kappa_{0})$. Indeed, when $\kappa_{1} \to \infty$, $\text{Disk}_{\kappa_{1}}(x,y; R)$ tends to zero as $1/(2\kappa_{1})$. In short, in the above abstract model $\kappa_{1}$ controls how each individual structure persists or decays in time and $\kappa_{0}$ how an individual structure meets and coexists with other emergent structures. Note that, although we have taken the center of the discs to be static, any disc can move with a constant velocity with $x$ and $y$ components $v_{x}$ and $v_{y}$ respectively, if we take $\text{Disk}_{\kappa_{1}}(x-v_{x}t,y-v_{y}t; R)$.

\subsection{Example 4: Embeddings having real or complex fractal objects as parts}

We now construct $\mathcal{B}_{\kappa}$ embeddings that connect a set of fractal objects at $\kappa=0$ to their ordinary sum as $\kappa \to \infty$. We first briefly sketch how these fractals are constructed (the reader is referred to \cite{CHAOSOLFRAC} for a detailed exposition of the method).

We first recall that any real number $x$ can be represented in radix $p \in \mathbb{N}$ as \cite{CHAOSOLFRAC}
\begin{equation}
x=\text{sign}(x)\sum_{k=-\infty}^{\left \lfloor \log_{p} |x| \right \rfloor}p^{k}\mathbf{d}_{p}(k,|x|) \label{realrad}
\end{equation}
where $\mathbf{d}_{p}(k,x)$ is the digit function defined in Eq. \ref{cucuAreal}. 

In \cite{CHAOSOLFRAC} we have shown how an arbitrary function $f(x)$ can be decomposed into any desired (finite) number of parts (fractal objects) whose ordinary addition is equal to $f(x)$ for every possible value of $x$ and regardless of the function $f(x)$ being nonlinear. Although the parts are discontinuous and non-differentiable, they possess all internal invariances of $f(x)$ and are constructed from this latter 'mother' function. Let $p>1$, $\lambda >1$ be natural numbers, let $n \in [0,\lambda-1]$ be an integer.

We define the $n$-th fractal object $\mathbf{F}_{p\lambda n}f(x)$ of a $p\lambda n$ fractal decomposition of $f(x)$ as
\begin{eqnarray}
\mathbf{F}_{p\lambda n}f(x)\equiv  \frac{2\ \text{sign}f(x)}{\lambda(\lambda-1)}\sum_{k=-\infty}^{\lfloor \log_{p}|f(x)| \rfloor}p^{k}\mathbf{d}_{p}(k,|f(x)|)\mathbf{d}_{\lambda}(0,|k|+n)  \label{lampm}
\end{eqnarray}

In \cite{CHAOSOLFRAC} we have proved that the fractal objects satisfy 
\begin{equation}
f(x)=\sum_{n=0}^{\lambda-1}\mathbf{F}_{p\lambda n}f(x) \label{theresult}
\end{equation}
i.e. their ordinary sum yields the function from which they constitute a decomposition. Indeed Eq. (\ref{theresult}) is equivalent to
\begin{equation}
f(x)=\text{sign}(f(x))\sum_{k=-\infty}^{\left \lfloor \log_{p} |f(x)| \right \rfloor}p^{k}\mathbf{d}_{p}(k,|f(x)|) \label{realradf}
\end{equation}
which is the expansion in radix $p$ of $f(x)$, as given by Eq. (\ref{realrad}).
Thus, these objects $\mathbf{F}_{p\lambda n}f(x)$ can be taken as parts at zero of an embedding that carries them to $f(x)$ at infinity
\begin{equation}
\mathcal{R}_{f(x)}(m,\kappa)=(1+2\kappa)\sum_{n=0}^{\lambda-1}\mathcal{B}_{\kappa}\left(m-n, \frac{1}{2}\right)\mathbf{F}_{p\lambda n}f(x) \label{theresultRKE}
\end{equation}

\begin{figure}
\includegraphics[width=1.0 \textwidth]{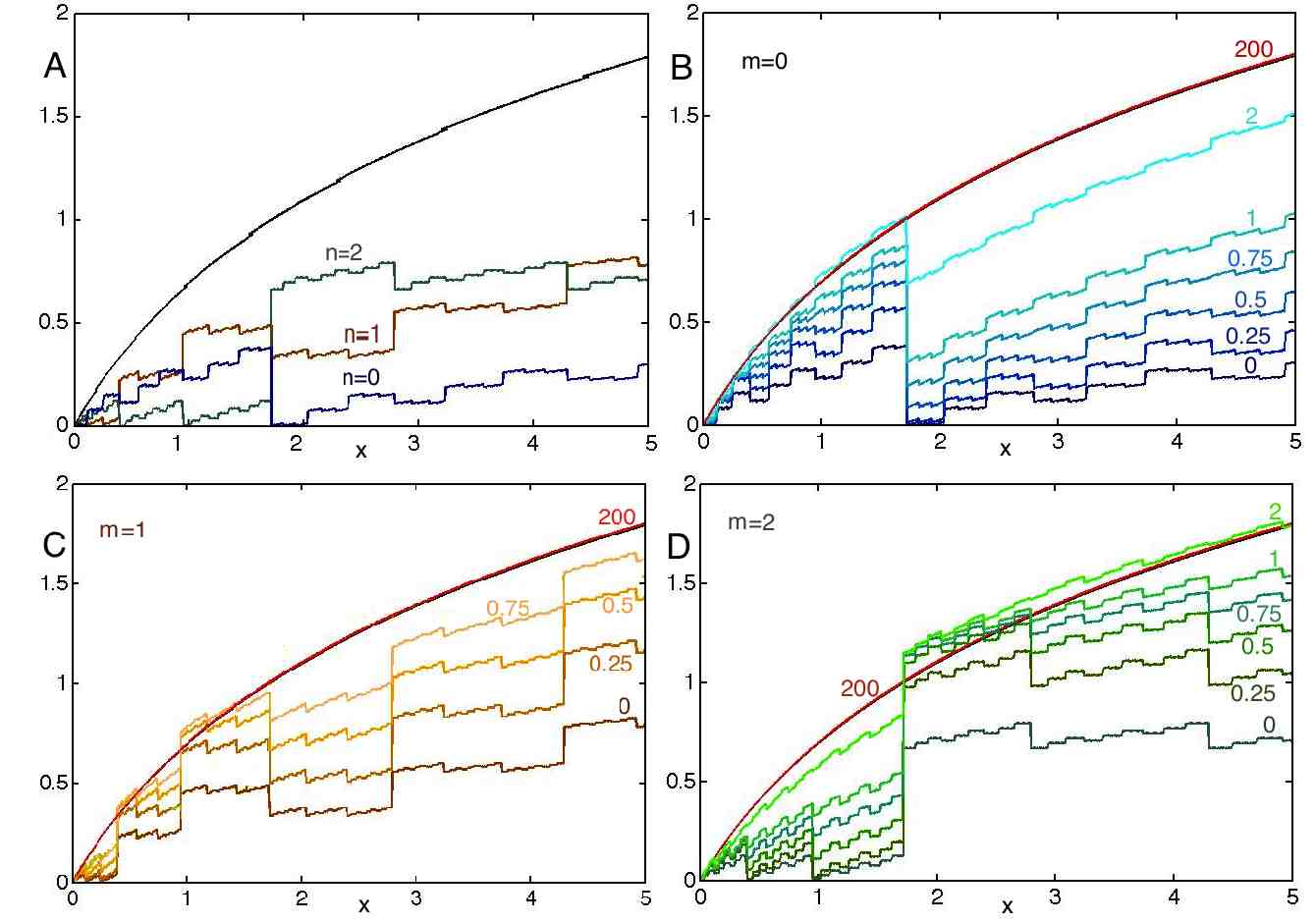}
\caption{\scriptsize{A) Plot of the function $f(x)=\ln (1+x)$ and the three fractal objects  $\mathbf{F}_{3,3,n}f(x)$ obtained from Eq. (\ref{lampm}) for the values of $n$ indicated. In panels B-D, the embedding $\mathcal{R}_{f(x)}(m,\kappa)$ obtained from Eq. (\ref{theresultRKE}) for $m=0$ (panel B), $m=1$ (panel B) and $m=2$ (panel C) is plotted for the fractal objects in panel A and the values of $\kappa$ indicated close to the curves.}} \label{logdec}
\end{figure}

\begin{figure}
\includegraphics[width=1.0 \textwidth]{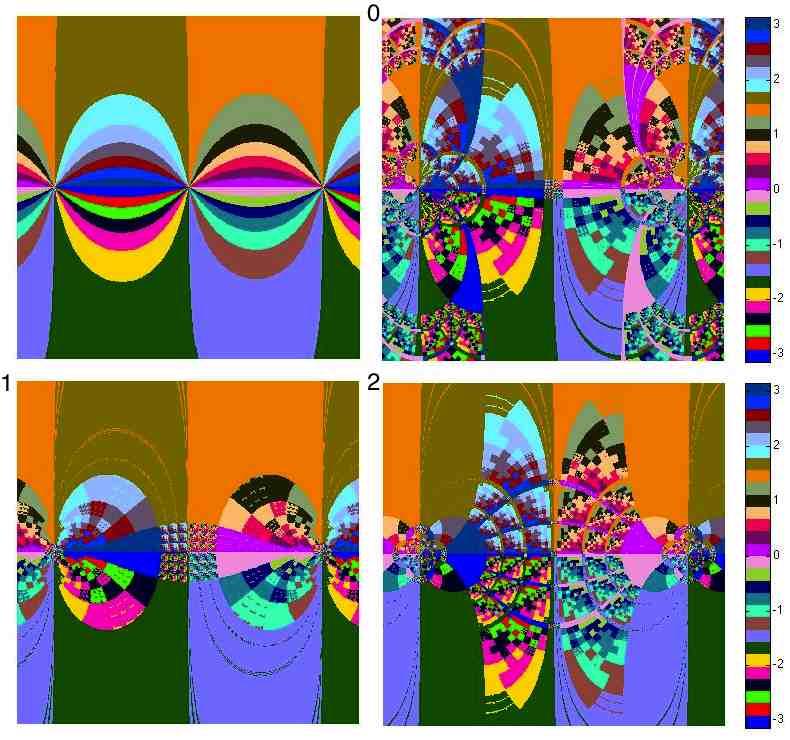}
\caption{\scriptsize{Phase plot of the function $f(z)=\tan z$ (top left panel) and the three fractal objects derived from it $\mathbf{F}_{3,3,n}\tan z$ from Eq. (\ref{lamcompl}) for the values of $n$ indicated on the panels. On each panel, shown is a square portion of the complex plane where the lower left corner is the point $z=-2-2i$ and the upper left corner the point $z=2+2i$.}} \label{decompl}
\end{figure}

In Fig. \ref{logdec}A, the function $f(x)=\ln (1+x)$ is plotted together with the three fractal objects $\mathbf{F}_{3,3,n}f(x)$ obtained taking $p=3$, $\lambda=3$ in Eq. (\ref{lampm}) for the values of $n$ indicated close to the curves in the figure. Inspection shows that the curve $f(x)$ lies above these fractal objects and that the sum of the latter is equal to $f(x)$ for every value of $x$. 
In Fig. \ref{logdec}B-D, the embedding Eq. (\ref{theresultRKE}) carrying each of the fractal objects (the corresponding value of $n$ is indicated in each panel) is plotted as a function of $\kappa$ (indicated close to the curves) and $x$. At $\kappa \sim 0$ each fractal object is as in Fig. \ref{logdec}A. As $\kappa$ increases, each of the fractal objects becomes gradually coarser and smoother and it approaches $f(x)$. At  $\kappa$ large, each fractal object  accurately approximates $f(x)$ for every value of $x$.

\begin{figure*}
\includegraphics[width=0.8 \textwidth]{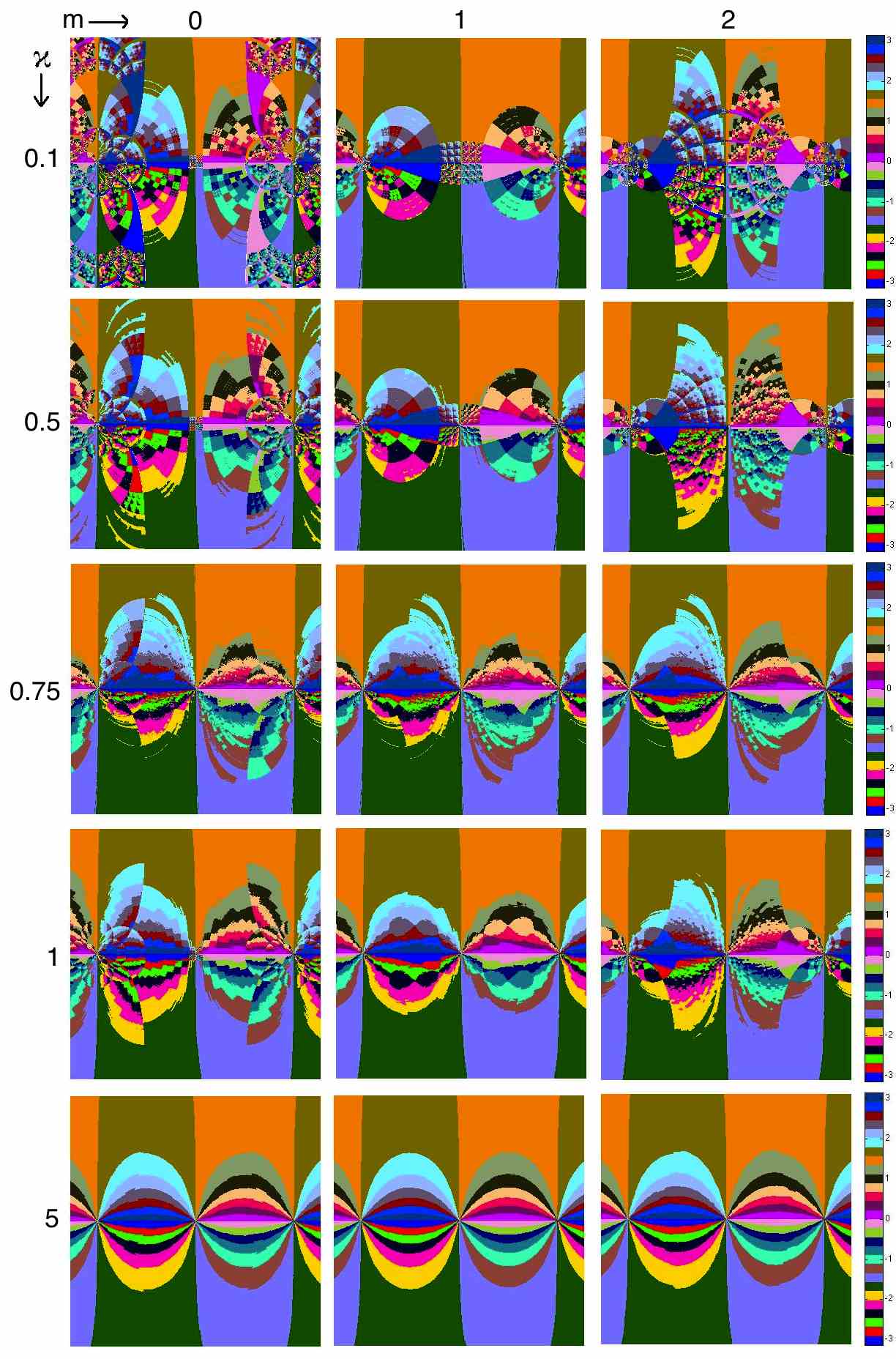}
\caption{\scriptsize{The embedding $\mathcal{R}_{f(z)}(m,\kappa)$ obtained from Eq. (\ref{theresultRKEc}) for the fractal objects $\mathbf{F}_{3,3,n}\tan z$ in Fig. \ref{decompl} as parts at zero, and the values of $m$ and $\kappa$ indicated in the figure. On each panel, shown is a square portion of the complex plane where the lower left corner is the point $z=-2-2i$ and the upper left corner the point $z=2+2i$.}} \label{decomplemb}
\end{figure*}

The above decomposition and the associated embedding can now be extended to the complex plane. 
We note that a complex number $z \in \mathbb{C}$ can be expanded as
\begin{equation}
z=\sum_{k=-\infty}^{\left \lfloor \log_{p} |z| \right \rfloor}p^{k}\mathbf{c}_{p}(k,z) \label{complrad}
\end{equation}
where we have introduced the complex digit function \cite{CHAOSOLFRAC}
\begin{equation}
\mathbf{c}_{p}(k,z)\equiv \text{sign}(x)\mathbf{d}_{p}(k,|x|)+i\ \text{sign}(y)\mathbf{d}_{p}(k,|y|)
\label{cdigit}
\end{equation}
with $x$ and $y$ being the real and the imaginary parts $z=x+iy$ of the complex variable $z$, respectively. 

The $p\lambda n$ fractal decomposition of a complex function $f(z)$, $f:\mathbb{C}\to \mathbb{C}$, yields $\lambda$ fractal objects on the complex plane $\mathbf{F}_{p\lambda n}f(z):\mathbb{C}\to \mathbb{C}$  and is given by \cite{CHAOSOLFRAC}
\begin{eqnarray}
&&\mathbf{F}_{p\lambda n}f(z) \equiv \frac{2}{\lambda(\lambda-1)}\sum_{k=-\infty}^{\lfloor \log_{p}|f(z)| \rfloor}p^{k}\mathbf{c}_{p}(k,f(z))\mathbf{d}_{\lambda}(0,|k|+n) \label{lamcompl}
\end{eqnarray}
and we have, as above
\begin{equation}
f(z)=\sum_{n=0}^{p-1}\mathbf{F}_{p\lambda n}f(z) \label{resulcom}
\end{equation}
Therefore, the embedding
\begin{equation}
\mathcal{R}_{f(z)}(m,\kappa)=(1+2\kappa)\sum_{n=0}^{\lambda-1}\mathcal{B}_{\kappa}\left(m-n, \frac{1}{2}\right)\mathbf{F}_{p\lambda n}f(z) \label{theresultRKEc}
\end{equation}
carries the fractal objects in the complex plane at zero to the `mother function' $f(z)$ at infinity.

In Fig. \ref{decompl} the phase plot of the function $f(z)=\tan z$ (top left panel) and of the three fractal objects derived from it $\mathbf{F}_{3,3,n}\tan z$ from Eq. (\ref{lamcompl}) are shown for the values of $n$ indicated on the panels. On each panel, a square portion of the complex plane is displayed where the lower left corner is the point $z=-2-2i$ and the upper left corner the point $z=2+2i$.

If we take these complex fractal objects at zero, we can investigate the behavior of Eq. (\ref{theresultRKEc}) on the complex plane. In Fig. \ref{decomplemb} the phase plot of the embedding $\mathcal{R}_{f(z)}(m,\kappa)$ obtained from Eq. (\ref{theresultRKEc}) for the fractal objects $\mathbf{F}_{3,3,n}\tan z$ in Fig. \ref{decompl} as parts at zero, and the values of $m$ and $\kappa$ indicated in the figure. On each panel, the same portion as in Fig. \ref{decompl} of the complex plane is shown. It is observed how, by increasing $\kappa$, each of the three fractal objects is gradually carried to $f(z)=\tan z$ at $\kappa=\infty$. Indeed, for $\kappa=5$ a good approximation of the function at infinity has already been achieved. It is to be noted that, as $\kappa$ increases more information of all fractal objects is incorporated to each branch $m$ of the embedding. At intermediate-large $\kappa$ values (e.g.) $\kappa=1$ the embedding is no longer a fractal on the complex plane, but the phase has complicated modulations compared to the function at infinity that gives the phase plot a certain  `wabi-sabi' \cite{Koren} quality.

\section{Application: Finding the roots of complex univariate polynomials}
\label{apli}

The problem of finding all roots of polynomials in the complex plane is not only a most important problem in numerical analysis, in particular, and mathematics, in general. Besides being crucial in many applications, it is also both intellectually and aesthetically rewarding \cite{Kalantari,Kalantari2,Gdawiec}. Powerful tools to deal with this problem are provided by the field of numerical algebraic geometry \cite{SW95}, which is concerned with the application and integration of  continuation methods to describe solution components of polynomial systems \cite{SVW, SVW04}. Standard numerical techniques, as continuation or path-following methods \cite{SVW, AG90a, AG90b, Mor87, Wat86, Wat89}, are used to trace the solution paths.

Let $a_{k} \in \mathbb{C}$, $k=\{0,1,\ldots, N\}$ be complex coefficients. A polynomial of a complex variable $z$ and degree $N \in \mathbb{N}$
\begin{equation}
P(z)=a_{N}z^{N}+a_{N-1}z^{N-1}+\ldots+a_{1}z+a_{0} \label{poly1}
\end{equation}
has $N$ (not necessarily distinct) roots $r_{n}$ in the complex plane, by the fundamental theorem of algebra. In terms of these roots, the solution of the algebraic equation $P(z)=0$ can equivalently be written as
\begin{equation}
(z-r_{N-1})(z-r_{N-2})\ldots(z-r_{0})=0 \label{poly2}
\end{equation}
Therefore, if one root, say $r_{m}$, is found, the monomial $(z-r_{m})$ can be factorized out of the polynomial $P(z)$ lowering its degree. Finding each individual root separately and factorizing it out are steps that belong to most numerical methods to obtain the roots of $P(z)$ as e.g. the celebrated Newton-Raphson method \cite{Bulirsch}. Here, we use $\mathcal{B}_{\kappa}$-embeddings to \emph{simultaneously} find all these roots without factorizing them out. Regarding numerical matters, our approach is here presented just as an straightforward application of $\mathcal{B}_{\kappa}$-embeddings and without any aim to compete with previous existing methods (see \cite{SVW} specially). Therefore, we refrain of e.g. discussing its performance and efficiency in comparison to those.

%Although this is something that is accomplished by other existing methods as well (see \cite{SVW} and references there) our method delivers some further valuable analytical (not numerical) information that may be used to investigate deep results of pure mathematics, in relationship with Galois theory (see the brief discussion at the end of this section). 

%We first note a result that can be used to quickly check whether if a numerical method \emph{has failed} to converge to the roots without having to replace each of them individually. The latter satisfy (among others) the following relationships \cite{Edwards}
%\begin{equation}
%\sum_{k=0}^{N-1}r_{k}=-\frac{a_{N-1}}{a_{N}} \qquad \qquad
%\prod_{k=0}^{N-1}r_{k}=(-1)^{N}\frac{a_{0}}{a_{N}} \label{conve} 
%\end{equation}
%as can be easily checked by comparing the coefficients accompanying the powers of $z^{0}$ and $z^{N-1}$ in Eqs. (\ref{poly1}) and (\ref{poly2}). Satisfying Eqs. (\ref{conve}) up to a certain tolerance $\epsilon$ is a necessary (but not sufficient) condition for the $r_{k}$'s to be the roots of $P(z)$, i.e.
%\begin{equation}
%\sum_{k=0}^{N-1}r_{k}=-\frac{a_{N-1}}{a_{N}} \qquad \qquad
%\prod_{k=0}^{N-1}r_{k}=(-1)^{N}\frac{a_{0}}{a_{N}} \label{conve} 
%\end{equation}

Our method proceeds as follows. We first note that from the structure of $P(z)$, an embedding can be constructed as
\begin{equation}
\mathcal{R}_{P(z)}(n,\kappa)=(1+2\kappa)\left[(a_{N}z^{N}+a_{0})\mathcal{B}_{\kappa}\left(n, \frac{1}{2}\right)+(a_{N-1}z^{N-1}+\ldots+a_{1}z)\mathcal{B}_{\kappa}\left(n-1, \frac{1}{2}\right)\right] \label{polyek1}
\end{equation} 
i.e., we split the polynomial $P(z)$ into two parts at zero, $P_{0}(z)=a_{N}z^{N}+a_{0}$ and $P_{1}(z)=a_{N-1}z^{N-1}+\ldots+a_{1}z$. At infinity, the embedding becomes equal to $P(z)$.
We first observe that the algebraic equation $P_{0}(z)=0$ involving the branch $n=0$ of the embedding can exactly be solved at zero to yield the $N$ roots on a regular $N$-gon
\begin{equation}
r_{0,m}=\left|\frac{a_{0}}{a_{N}}\right|^{1/N}e^{\frac{i}{N}\left(\text{Arg}(-a_{0}/a_{N})+2\pi m \right)} \qquad m=0,1,\ldots, N-1 \label{cyclotom}
\end{equation}
which is inscribed on a circumference of radius $\left|a_{0}/a_{N}\right|^{1/N}$ in the complex plane. We now note that, for $\kappa >0$, the embedding $\mathcal{R}_{P(z)}(n,\kappa)$ is always a polynomial of degree $N$ in $z$, which means that, for every nonvanishing value of $\kappa$
there are always $N$ roots in the complex plane. Since Eq. (\ref{cyclotom}) provides the $N$ roots at $\kappa=0$ on the branch $n=0$ the strategy is, by following that branch from zero to infinity, to follow the gradual change of location of the roots given by Eq. (\ref{cyclotom}) as $\kappa$ increases, by exploiting the analyticity of, both, the polynomial $P(z)$ and the embedding $\mathcal{R}_{P(z)}(n,\kappa)$. In this way, we reformulate the problem of finding the roots of $P(z)$ as the problem of finding the values of $z$ such that, for any specific value of $\kappa$,
\begin{equation}
\mathcal{R}_{P(z)}(0,\kappa)=0 \label{thequa}
\end{equation} 
by knowing that the equation
\begin{equation}
\mathcal{R}_{P(z)}(0,0)=0 \label{thequa0}
\end{equation} 
has the $N$ solutions given by Eq. (\ref{cyclotom}).

%We now note that if we consider $z=r_{\kappa,m}$ and, hence, the vector $\mathbf{z}=\mathbf{r}_{\kappa}$ we can now obviously write Eq. (\ref{thequa}) in vector form \emph{for every value of $\kappa$} as
%\begin{equation}
%\mathcal{R}_{\mathbf{P}(\mathbf{r}_{\kappa})} (0,\kappa)=\mathbf{0} \label{thequave}
%\end{equation} 
%where it is now understood that the embedding acts componentwise (and, hence, is itself a vector). 

\emph{Because of the analyticity of both $P(z)$ and the embedding}, each of the roots $r_{0,m}$ in Eq. (\ref{cyclotom}) becomes displaced to $r_{\kappa,m}$ through the $\kappa$ deformation. The latter are the roots of the embedding at non-vanishing $\kappa$. Thus if we take $z=r_{\kappa,m}$ we, obviously, have
\begin{equation}
\mathcal{R}_{P(r_{\kappa,m})} (0,\kappa)=0 \qquad m=0,\ldots, N-1 \label{thequave}
\end{equation} 

By differentiating vs. $\kappa$ this latter equation, we obtain
\begin{equation}
\frac{d\mathcal{R}_{P(r_{\kappa,m})}(0,\kappa)}{d\kappa}=\frac{\partial \mathcal{R}_{P(r_{\kappa,m})}(0,\kappa)  }{\partial \kappa}+\frac{\partial \mathcal{R}_{P(r_{\kappa,m})}(0,\kappa)  }{\partial r_{\kappa,m}} \frac{dr_{\kappa,m}}{d\kappa}=0
\end{equation}
from which we find the flow
\begin{equation}
\frac{dr_{\kappa,m}}{d\kappa}=-\frac{\partial \mathcal{R}_{P(r_{\kappa,m})}}{\partial \kappa} \left[ \frac{\partial \mathcal{R}_{P(r_{\kappa,m})}  }{\partial r_{\kappa,m}}       \right]^{-1} 
\qquad m=0,\ldots, N-1 \label{theflowdir}
\end{equation}
where
\begin{eqnarray}
\frac{\partial\mathcal{R}_{P(r_{\kappa,m})}}{\partial \kappa}&\equiv&(1+2\kappa)\sum_{j=0}^{1}P_{j}(r_{\kappa,m})\left(\frac{2\mathcal{B}_{\kappa}\left(j,\frac{1}{2} \right)}{1+2\kappa}+\frac{d}{d\kappa}\mathcal{B}_{\kappa}\left(j,\frac{1}{2} \right) \right) \label{Watson} \\
\frac{\partial \mathcal{R}_{P(r_{\kappa,m})}  }{\partial r_{\kappa,m}}&\equiv&(1+2\kappa)\sum_{j=0}^{1}\frac{dP_{j}(r_{\kappa,m})}{dr_{\kappa,m}}\mathcal{B}_{\kappa}\left(j,\frac{1}{2} \right)  \label{naca}\\
\frac{d}{d\kappa}\mathcal{B}_{\kappa}\left(j,\frac{1}{2} \right)&=&
\frac{1}{2\kappa^{2}}\left[\left(j+\frac{1}{2}\right)\tanh^{2}\left(\frac{j+\frac{1}{2}}{\kappa} \right)-\left(j-\frac{1}{2}\right)\tanh^{2}\left(\frac{j-\frac{1}{2}}{\kappa} \right)-1
\right]  \\
P_{0}(r_{\kappa,m})&=&a_{N}r_{\kappa,m}^{N}+a_{0} \qquad  P_{1}(r_{\kappa,m})=a_{N-1}r_{\kappa,m}^{N-1}+a_{N-2}r_{\kappa,m}^{N-2}+\ldots+a_{1}r_{\kappa,m} \\
\frac{dP_{0}(r_{\kappa,m})}{dr_{\kappa,m}}&=&Na_{N}r_{\kappa,m}^{N-1} \qquad \quad \ \ \frac{dP_{1}(r_{\kappa,m})}{dr_{\kappa,m}}=(N-1)a_{N-1}r_{\kappa,m}^{N-2}+\ldots+a_{1} \label{nacar}
\end{eqnarray}

We discretize Eq. (\ref{theflowdir}) by means of Euler's method, selecting a small but finite $\Delta \kappa$ increment. We finally obtain
\begin{equation}
r_{\kappa+\Delta\kappa,m}=r_{\kappa,m}-\Delta \kappa \frac{\partial \mathcal{R}_{P(r_{\kappa,m})}}{\partial \kappa} \left[ \frac{\partial \mathcal{R}_{P(r_{\kappa,m})}  }{\partial r_{\kappa,m}}       \right]^{-1} \label{theflow}
\end{equation}
Note that all $N$ equations are decoupled and, hence, can be computed in parallel.
To this goal it is useful to define the $N$-component vector $\mathbf{r}_{\kappa}$, containing the roots of Eq. (\ref{thequa}). At $\kappa=0$ we thus have $\mathbf{r}_{0}=(r_{0,0}, r_{0,1}, \ldots, r_{0,N-1})$, where the $r_{0,m}$ are all given by Eq. (\ref{cyclotom}). At any other value of $\kappa$, each component of $\mathbf{r}_{\kappa}$ is given by iterating Eq. (\ref{theflow}) up to that $\kappa$ value. We are thus able to simultaneously find all roots of $P(z)$ at $\kappa$ finite (sufficiently large) to any degree of accuracy: The roots are the components of the vector $\mathbf{r}_{\infty}$.

We note that, asymptotically, for $\kappa$ large
\begin{eqnarray}
\frac{\partial\mathcal{R}_{P(r_{\kappa,m})}}{\partial \kappa} &\sim& \frac{P(r_{\kappa,m})}{\Delta \kappa} \\
\frac{\partial \mathcal{R}_{P(r_{\kappa,m})}  }{\partial r_{\kappa,m}}  &\sim& \frac{dP_{j}(r_{\kappa,m})}{dr_{\kappa,m}} = P'(r_{\kappa,m})
\end{eqnarray}
Therefore, we have that Eq. (\ref{theflow}) becomes, in this regime
\begin{equation}
r_{\kappa+1,m}=r_{\kappa,m}-\frac{P(r_{\kappa,m})}{P'(r_{\kappa,m})} \qquad m=0,\ldots, N-1 \label{NR}
\end{equation}
which is the Newton-Raphson iteration \cite{Bulirsch}. Thus, asymptotically, in the infinite limit of the $\mathcal{B}_{\kappa}$-embedding, which coincides with the polynomial $P(z)$ whose roots we want to find, our method becomes Newton's method for initial conditions \emph{that are already excellent approximations to the roots}. Thus, our \emph{global} method gradually deforms to a \emph{local} method (the Newton-Raphson method) when the roots are nearly approached. Note that \emph{no higher derivative other than the first-order one is required at any iteration step}. Furthermore, with our method: 1) \emph{we do not need to make any good initial guess} as in the Newton-Raphson method: The initial input is calculated from Eq. (\ref{cyclotom}) \emph{for any arbitrary polynomial}; 2) \emph{we do not need to deflate the polynomial} \cite{Bulirsch} once we have found a specific root because all roots are simultaneously obtained with the right multiplicity and contained in the vector $\mathbf{r}_{\infty}$; furthermore, and importantly, 3) because of the saturation properties of the $\mathcal{B}_{\kappa}$-function, we can be confident that for $5< \kappa <10$ the roots are already well approximated and, therefore, our Eq. (\ref{theflow}) can be safely switched to Eq. (\ref{NR}) (if we want to polish the roots \emph{quadratically} to a large number of significant digits). 

\begin{figure*}
\includegraphics[width=1 \textwidth]{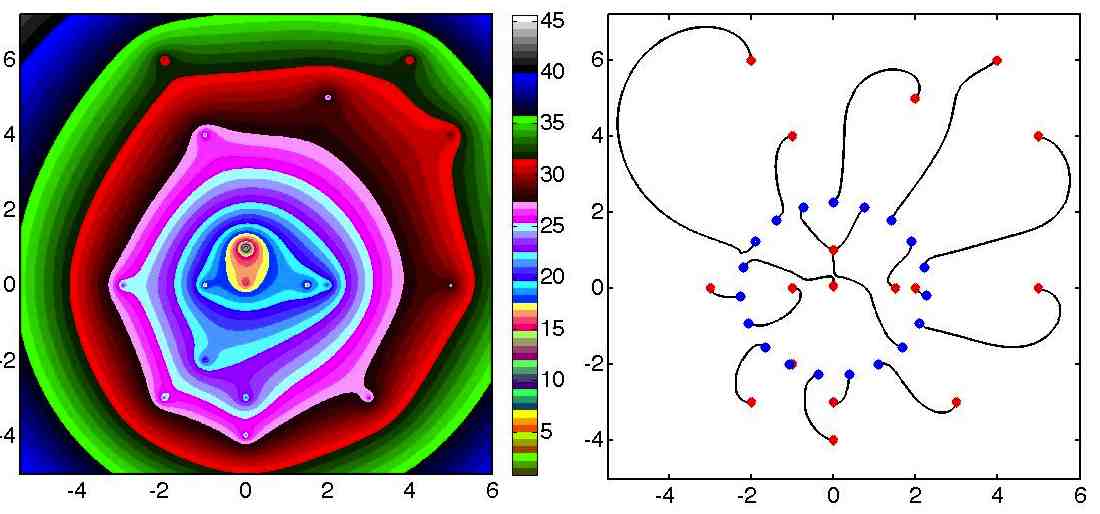}
\caption{\scriptsize{(Left) Plot of the function $\ln (1+|P(z)|)$ in the complex plane for a polynomial of 19th-degree, with roots at $z=1.5$, $2$, $3-3i$, $5+4i$, $5$, $-1$, $-1-2i$, $-3i$, $-4i$, $0.05i$, $-2+6i$, $-3$, $-1+4i$, $2+5i$, $4+6i$, $-2-3i$ and $i$ (this latter root with multiplicity equal to 3). The location of the roots is clearly seen as isolated points surrounded of little circles of a different color than the surrounding region. (Right) Evolution in the complex plane of the flow given by Eq. (\ref{theflow}) for the polynomial $P(z)$ of the left panel, for an initial condition given by Eq. (\ref{cyclotom}) and consisting of the location of the vertices of a regular 19th-gon (blue circles). The flow correctly converges to the roots (red circles) no matter how far away these are from the initial condition, following the continuous paths in the complex plane shown in the figure.}} \label{polypot}
\end{figure*}

In Fig. \ref{polypot} (left), the modulus of a 19th-degree polynomial $P(z)$ is shown in the complex plane. In order to render the details appropriately (since the modulus varies from $0$ to $\approx e^{45}$) the function $\ln (1+|P(z)|)$ is plotted for the region in the complex plane where the roots are located. The latter are clearly visible as 17 isolated points, the root at $+i$ having multiplicity equal to 3 (for the values of all the roots see the figure caption). In Fig. \ref{polypot} (right) the result obtained with use of our method is shown. Starting at $\kappa=0$ from the vertices of a regular $19$-gon whose positions in the complex plane are calculated from Eq. (\ref{cyclotom}) and which are shown in the figure as blue dots, by iterating Eq. (\ref{theflow}) for a fixed increment $\Delta \kappa=0.003$, the roots trace each of the paths shown in the figure. At infinity, these paths converge to the roots of $P(z)$ that we wanted to find (red circles), all being cleanly found with their right multiplicity.

\begin{figure*}
\includegraphics[width=1.0 \textwidth]{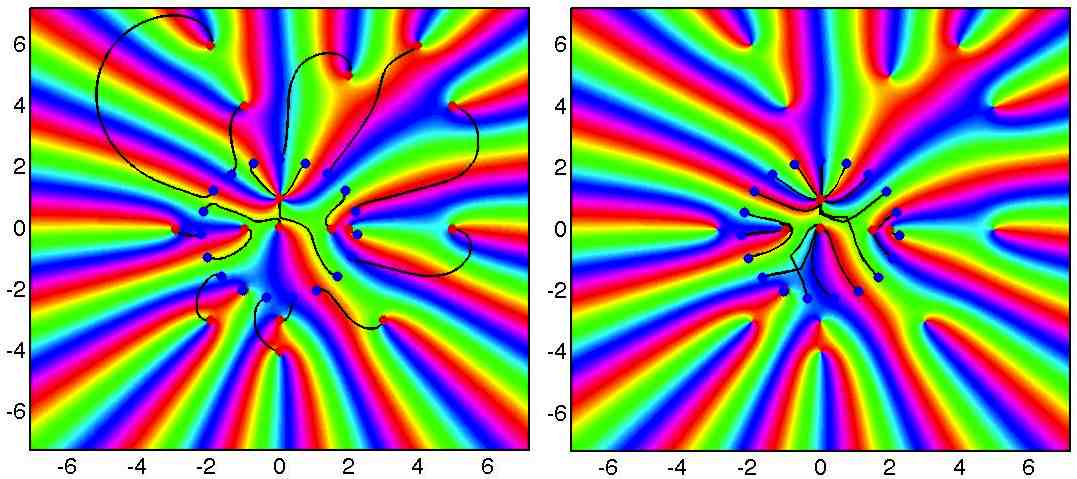}
\caption{\scriptsize{(Left) Phase plot in the complex plane of $P(z)$ in Fig. \ref{polypot} on which our result, in Fig. \ref{polypot} (right) is also superimposed. Our method correctly finds all the roots of the polynomial with their right multiplicity. (Right) Result obtained by means of the Newton-Raphson method, Eq. (\ref{NR}) for the same initial conditions. The latter fails to find the roots outside the initial 19-gon because the paths tend always to follow isochromatic lines in the direction of local lower modulus.}} \label{newr}
\end{figure*}

As indicated above, the Newton-Raphson is fine to polish the roots once we are near to them. 
In Fig. \ref{newr} (left) we plot the phase of $P(z)$ \cite{Wegert} by superimposing our result as well (i.e. the right panel of Fig. \ref{polypot}), and we see that the paths obtained from our method, which contains global information thanks to the embedding, has no problem in crossing separatrices and going against the direction of local lower modulus. Quite on the contrary, Newton's method, being local, tends to follow isochromatic lines (constant phase) downhill. As a result, Newton's method does not detect the roots outside the $19$-gon in whose vertices the initial condition is located: The paths all collapse to the roots inside the $19$-gon and, therefore, the multiplicity with which the roots are found is also incorrect.

It is to be noted that our method replaces the problem of looking for the solutions of $P(z)$ by the problem of finding the connected loci that satisfy Eq. (\ref{thequa}) for all $\kappa$ positive. We call the set of such loci the zero $\kappa$-net. This problem is simpler than the original one since we know that the point $\mathbf{r}_{0}$ whose $N$ coordinates are given by Eq. (\ref{cyclotom}) belongs to the set and we can use that the $\mathcal{B}_{\kappa}$-function is continuous and differentiable, besides its nice saturation properties, to get to the solution through a continuous flow.

We now discuss and fix certain convergence problems that our method may experience. As simply stated above, our method should fail to converge in certain exceptional cases where the roots are aligned in such a way that the paths obtained from the flow meet the line containing the roots orthogonally in a pure symmetric manner. In such scenario the zero $\kappa$-net has points of self-intersection  where the individuality of the branches of the $\kappa$-net (as numerically obtained by the differential flow) is compromised. The larger the degree of the polynomial, the more rare this situation is to happen. This problem can be fixed once and for all, in general, in an easy way, as we discuss below, by exploiting the freedom that we have to choose embeddings with the same parts at zero and at infinity.
 
\begin{figure}
\includegraphics[width=1.0\textwidth]{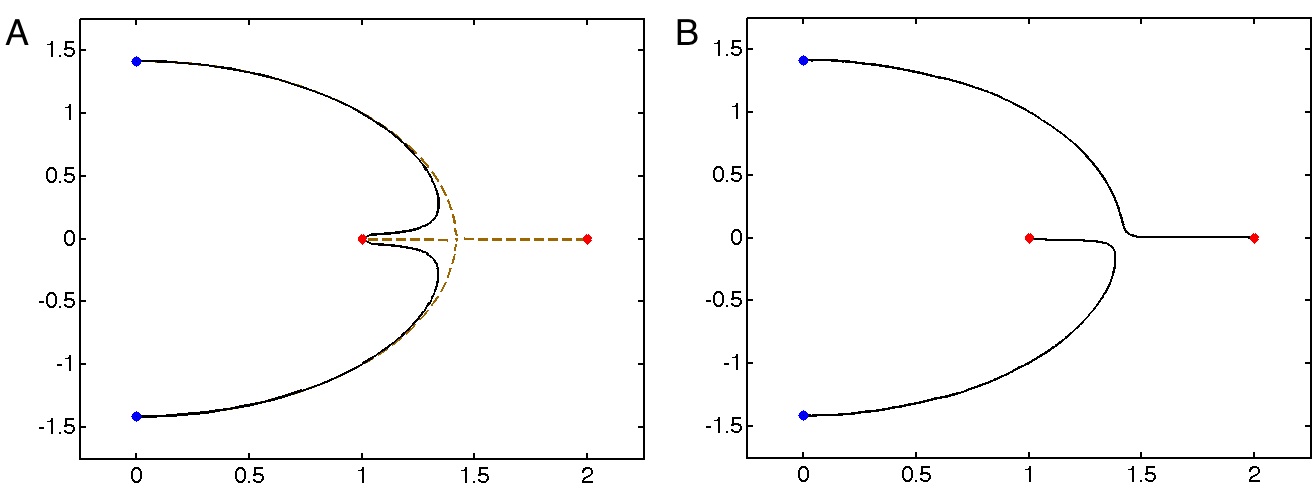}
\caption{\scriptsize{A. Plot in the complex plane of the zero $\kappa$-net (discontinuous curves) obtained from Eq. (\ref{knet}) and the paths that result from iterations of Eq. (\ref{theflow}) (continuous curves) for the polynomial $P(z)=z^{2}-3z+2$. The method fails to locate the root at $z=2$. B. Plot in the complex plane of the paths that result from iterations of Eq. (\ref{theflow2}). The method successfully finds both two roots $r_{\infty,+}=2$, $r_{\infty,-}=1$ providing a good approximation of the zero $\kappa$-net as well. In both cases $\Delta \kappa=0.003$.}} \label{sybre}
\end{figure}

Quite fortunately, the way to proceed with these exceptional cases (and to understand and fix the problem, in general) can be elucidated in detail for a simplest case in which the problem is already met (and where the zero $\kappa$-net can exactly be solved): The case of quadratic polynomials. For the algebraic equation
\begin{equation}
a_{2}z^{2}+a_{1}z+a_{0}=0
\end{equation}
the zero $\kappa$-net has two branches that are exactly obtained from Eq. (\ref{thequa}) as 
\begin{equation}
r_{\kappa,\pm}=\frac{-a_{1}(\kappa)\pm \sqrt{a_{1}^{2}(\kappa)-4a_{2}(\kappa)a_{0}(\kappa)}}{2a_{2}(\kappa)} \label{knet}
\end{equation} 
where
\begin{eqnarray}
a_{0}(\kappa)&=&(1+2\kappa)a_{0}\mathcal{B}_{\kappa}\left(0, \frac{1}{2}\right) \\ 
a_{1}(\kappa)&=&(1+2\kappa)a_{1}\mathcal{B}_{\kappa}\left(1, \frac{1}{2}\right) \\
a_{2}(\kappa)&=&(1+2\kappa)a_{2}\mathcal{B}_{\kappa}\left(0, \frac{1}{2}\right)
\end{eqnarray}
At the critical value $\kappa^{*}$ where $a_{1}^{2}(\kappa^{*})-4a_{2}(\kappa^{*})a_{0}(\kappa^{*})=0$ the zero $\kappa$-net has a point of self-intersection in which the two branches collide. Since the imaginary part of the branches vanishes, the different signs accompanying these branches are also lost and the numerical method fails to distinguish the branches in subsequent iterations. In Fig. \ref{sybre}A we show the exact solution of the zero $\kappa$-net for the polynomial $P(z)=z^{2}-3z+2$ as obtained from Eq. (\ref{knet}) (discontinuous curves) and the solution obtained iterating Eq. (\ref{theflow}) (continuous curves) which fails to locate the real root at $z=2$.

This problem is only caused by the fully symmetric arrangement of the trajectories of the iteration method with respect to the real axis that leads the two branches to come close to a self-intersection point of the zero $\kappa$-net. The solution to achieve convergence, in general, is to introduce a small, random perturbation at each iteration step, while always keeping within an embedding of the same class as the one with the unperturbed iterations. We have considerable freedom to do this, and the simplest way to achieve our goal (and one that fixes the problem once and for all) is to make the transformations $P_{0}(r_{\kappa,m}) \to P_{0}(r_{\kappa,m})+e^{i2\pi \theta_{\kappa}}$ and $P_{1}(r_{\kappa,m}) \to P_{1}(r_{\kappa,m})-e^{i2\pi \theta_{\kappa}}$ at each $\kappa \ne 0$. Here $\theta_{\kappa} \in \mathbb{R}$ is a random number drawn from the uniform distribution in the unit interval at each $\kappa \ne 0$. Note that neither the parts at zero nor the part at infinity change, and by adding the noisy term $(-1)^{j}e^{i2\pi \theta_{\kappa}}$ we preserve at each iteration step the distinctness of the branches through introducing a perturbation that averages out to zero, \emph{only at $\kappa \to 0$ and $\kappa \to \infty$}, softly breaking the symmetry. Thus, the path belongs to the same class of the original embedding and any harmful underlying symmetries as the one described above are softly broken. This suffices to warrant convergence of the method in general. The resulting solution paths being also a good numerical approximation of the zero $\kappa$-net for every $\kappa$ value.

We thus consider iterations of the map 
\begin{equation}
r_{\kappa+\Delta\kappa,m}=r_{\kappa,m}-\Delta \kappa \frac{\partial \widetilde{\mathcal{R}}_{P(r_{\kappa,m})}}{\partial \kappa} \left[ \frac{\partial \mathcal{R}_{P(r_{\kappa,m})}  }{\partial r_{\kappa,m}}       \right]^{-1} \qquad m=0,\ldots, N-1 \label{theflow2}
\end{equation}
where
\begin{equation}
\frac{\partial\widetilde{\mathcal{R}}_{P(r_{\kappa,m})}}{\partial \kappa}\equiv (1+2\kappa)\sum_{j=0}^{1}\left(P_{j}(r_{\kappa,m})+(-1)^{j}e^{i2\pi \theta_{\kappa}}\right)\left(\frac{2\mathcal{B}_{\kappa}\left(j,\frac{1}{2} \right)}{1+2\kappa}+\frac{d}{d\kappa}\mathcal{B}_{\kappa}\left(j,\frac{1}{2} \right) \right) \label{Watson2}
\end{equation} 
all other Eqs. (\ref{naca}) to (\ref{nacar}) remaining unaffected. The map now contains a class-preserving random perturbation at each step. Note that although the phase $\theta_{\kappa}$ changes randomly at each iteration step, this change is independent of $z$ (and, hence, the same for every $m$) and, therefore, it does not affect the derivative  $\frac{dP_{j}(r_{\kappa,m})}{dr_{\kappa,m}}$ in Eq. (\ref{naca}).

By iterating Eq. (\ref{theflow2}) for the polynomial $P(z)=z^{2}-3z+2$ where Eq. (\ref{theflow}) failed the method successfully finds now both two roots $r_{\infty,+}=2$, $r_{\infty,-}=1$ as seen in Fig. \ref{sybre}B. The obstacle at the point of self-intersection of the zero $\kappa$-net is swiftly overcome, the distinctness of the branches being preserved. The perturbation is tiny everywhere and does not compromise the smoothness of the flow. Only close to the point of self-intersection does the perturbation play a more prominent role, precluding that the distinctness of the branches is lost and keeping the flow smooth.

\section{Conclusions}

In this work we have presented the theory of $\mathcal{B}_{\kappa}$-embeddings illustrating it with a number of explicit specific examples which suggest ways in which $\mathcal{B}_{\kappa}$-embeddings may be useful to model irreversible processes involving nonlinear physical systems. As we have shown, $\mathcal{B}_{\kappa}$ embeddings allow any finite or denumerable set of objects (parts) to merge together with their ordinary sum as the embedding parameter $\kappa$ ranges from $0$ to $\infty$. We have applied these embeddings to fractal objects obtained out of a $p\lambda n$ fractal decomposition \cite{CHAOSOLFRAC}.

We have also provided an application to a problem of utmost interest in numerical analysis, by devising a method to find all roots of complex univariate polynomials. Certain convergence issues have been addressed and fixed and the method is robust and general, being applicable to univariate polynomials of any degree. Our method treats the zeros of the polynomial in the complex plane as analogous to electrical charges. By virtue of the analiticity of any polynomial, our method establishes a differential flow that behaves as a continuity equation in the complex plane in which the number of zeros is conserved and which continuously carries the zeros from a known solution of a known polynomial to the solutions of the unknown polynomial that we want to find. The method only needs the first derivative of the polynomial in question and may be extended to systems of nonlinear equations. 

The mathematics behind $\mathcal{B}_{\kappa}$-embeddings is quite elementary, yet this concept provides a unified perspective to model nonlinear complex systems, both discrete and continuous, that can exhibit a daunting variety of complex behavior in their spatiotemporal evolution. This is so because 
the superposition principle of linear systems can be generalized to nonlinear systems \cite{CHAOSOLFRAC} and $\mathcal{B}_{\kappa}$-embeddings show how any finite number of  independent nonlinear objects can be continuously superimposed so as to be embedded and merge together with their ordinary sum as the embedding parameter $\kappa$ is varied.

\bibliography{biblos}{}
\bibliographystyle{h-physrev3.bst}

\end{document}